\documentclass[final,12pt]{elsarticle}

\usepackage[margin=2.5cm]{geometry}

\usepackage{amssymb}
\usepackage{amsmath}

\usepackage{lineno}

\usepackage{xurl}

\newcounter{bla}

\journal{arXiv}

\begin{document}

\begin{frontmatter}



\title{uniGasFoam: a particle-based OpenFOAM solver for multiscale rarefied gas flows}


\author[a]{N. Vasileiadis\corref{author}}
\author[a]{G. Tatsios}
\author[b]{C. White}
\author[c]{D. A. Lockerby}
\author[a]{M. K. Borg}
\author[a]{L. Gibelli}

\cortext[author] {Corresponding author.\\\textit{E-mail address: } nvasilei@ed.ac.uk}
\address[a]{School of Engineering, Institute of Multiscale Thermofluids, University of Edinburgh, Edinburgh EH9 3FB, United Kingdom}
\address[b]{James Watt School of Engineering, University of Glasgow, Glasgow G12 8QQ, United Kingdom}
\address[c]{School of Engineering, University of Warwick, Coventry CV4 7AL, United Kingdom}

\begin{abstract}

This paper presents uniGasFoam, an open-source particle-based solver for multiscale rarefied gas flow simulations, which has been developed within the well-established OpenFOAM framework, and is an extension of the direct simulation Monte Carlo (DSMC) solver dsmcFoam+. The developed solver addresses the coupling challenges inherent in hybrid continuum-particle methods, originating from the disparate nature of finite-volume (FV) solvers found in computational fluid dynamics (CFD) software and DSMC particle solvers. This is achieved by employing alternative stochastic particle methods, resembling DSMC, to tackle the continuum limit. The uniGasFoam particle-particle coupling produces a numerical implementation that is simpler and more robust, faster in many steady-state flows, and more scalable for transient flows compared to conventional continuum-particle coupling. The presented framework is unified and generic, and can couple DSMC with stochastic particle (SP) and unified stochastic particle (USP) methods, or be employed for pure DSMC, SP, and USP gas simulations. To enhance user experience, optimise computational resources and minimise user error, advanced adaptive algorithms such as transient adaptive sub-cells, non-uniform cell weighting, and adaptive global time stepping have been integrated into uniGasFoam. In this paper, the hybrid USP-DSMC module of uniGasFoam is rigorously validated through multiple benchmark cases, consistently showing excellent agreement with pure DSMC, hybrid CFD-DSMC, and literature results. Notably, uniGasFoam achieves significant computational gains compared to pure dsmcFoam+ simulations, rendering it a robust computational tool well-suited for addressing multiscale rarefied gas flows of engineering importance.

\end{abstract}


\end{frontmatter}





\section{Introduction} \label{Sect:intro}

Multiscale phenomena are critical in several applications of major technological importance, such as micro-electro-mechanical and micropropulsion systems \citep{Aktas2002,Yang2017}, planetary atmosphere entry \citep{Wang2003B} and hypersonic flight \citep{Anderson2006}. These applications span a wide range of gas densities and length scales and it is increasingly common for the full spectrum of gas rarefaction ranging from the continuum to the free-molecular regime to be present and required to be modelled. The degree of gas rarefaction is measured by the Knudsen number $ \mathrm{Kn} =\lambda/L$, which is the ratio of the gas mean free path $\lambda$ to a characteristic length scale $L$. According to the Knudsen number, the flow regimes are commonly classified as continuum ($\mathrm{Kn}<0.001$), slip ($0.001<\mathrm{Kn}<0.1$), transition ($0.1<\mathrm{Kn}<10$), and free-molecular ($\mathrm{Kn}>10$) \citep{Karniadakis2005}.

In the continuum and slip flow regimes, the Navier-Stokes-Fourier (NSF) equations with suitable slip boundary conditions \citep{Maxwell1878,Smoluchowski1898} provide highly accurate predictions with low computational cost. However, in the transition and free-molecular regimes, reduced rates of intermolecular interactions lead to the breakdown of the linear constitutive laws, and thus the NSF equations become invalid.

In contrast, the Boltzmann equation, which is based on the kinetic theory of gases, has been proven to be accurate in the whole range of gas rarefaction \citep{Cercignani1988}. However, the solution of the Boltzmann equation poses a significant computational challenge compared to the NSF equations, and thus considerable effort has been devoted to the development of efficient numerical solution methods. Among these methods, the direct simulation Monte Carlo (DSMC) stands out as the most widely used \citep{Bird1994}. DSMC is a stochastic particle-based method that emulates the physics of the Boltzmann equation by tracking the motions and collisions of a large number of computational particles, providing simulation results that converge to the solution of the Boltzmann equation as the number of particles approaches infinity \citep{Wagner1992}. More specifically, in DSMC, the particle collision and streaming dynamics are split over a small time step and a computational grid is introduced to resolve intermolecular collisions and calculate macroscopic quantities by averaging particle properties. Although in principle DSMC provides an accurate description of gas flow over the full range of gas rarefaction, the cell size and time step are limited by the local mean free path and mean collision time respectively. Thus, the continuum regime becomes computationally intractable to DSMC since it requires a large number of computational cells and particles, as well as, a small time step to accurately resolve the small mean free path and mean collision time that characterize the continuum limit.

Over the past few decades, significant efforts have been made to address the need to accurately and efficiently simulate multiscale rarefied gas flows, as no single computational method currently achieves this. This has led to the development of hybrid continuum-DSMC methods, where finite volume (FV) computational fluid dynamics (CFD) and DSMC are coupled together, each resolving the continuum and rarefied flow regions, respectively. The most common hybrid approaches are based on the state-based domain decomposition (DD) \citep{Wadsworth1990, Roveda1998} splitting the domain into DSMC/CFD regions with simple hand-shaking regions for setting boundary conditions and using iterations to achieve a converged solution for steady-state cases. The heterogeneous multiscale method (HM) \citep{Garcia1999,Docherty2014,Lockerby2015,Docherty2016} offers a different route towards correcting the NSF equations in the rarefied regimes. In the HM method, localized DSMC regions are implemented throughout the flow domain to extract corrections for the NSF linear constitutive laws and boundary velocity slip and temperature jump coefficients. Several hybrid CFD-DSMC general-purpose solvers have been developed based on the DD approach, such as hy2Foam \citep{Espinoza2016}, HALO3D \citep{Vagishwari2022, Casseau2022}, and hybridDCFoam \citep{Vasileiadis2024}.

The disparate nature of the finite-volume and particle solvers in CFD-DSMC hybrids has revealed many numerical challenges. One significant drawback is the requirement for data exchange between the CFD and DSMC solvers. More specifically, the macroscopic quantities, such as pressure, temperature, and velocity, are computed by averaging the particle properties in the DSMC cells adjacent to the CFD region boundary. These quantities are then imposed as Dirichlet boundary conditions in the CFD solver. Similarly, the macroscopic quantities computed by the CFD solver are used to generate particles at the DSMC region boundaries. This coupling methodology can become challenging and costly if the solvers are not designed within the same computational framework. When solvers operate within different frameworks, challenges such as data compatibility, communication overhead, and the necessity for extensive code modifications may arise. These issues can increase the complexity of the coupling process and adversely affect overall performance. In addition, both coupled solvers should be open-source to enable developers to freely integrate different methodologies and algorithms seamlessly. Another challenge is that CFD-DSMC hybrids are not inherently robust, since noisy DSMC data passed to the CFD solver can cause instabilities. Thus CFD-DSMC hybrids require an experienced user to run both smoothly and accurately. Furthermore, on the one hand, in steady-state cases, CFD-DSMC hybrids require multiple hybrid iterations to converge, which can sometimes negate the potential computational savings of the hybrid method. On the other hand, in time-dependent scenarios, the CFD and DSMC solvers are coupled at regular intervals, presenting two primary challenges. First, the coupling between the CFD and DSMC solvers must be executed with high efficiency to prevent it from becoming the computational bottleneck. Second, a substantial number of DSMC particles are necessary to minimize statistical noise and prevent instabilities in the CFD solver, thereby reducing the scalability of the CFD-DSMC hybrid approach.

In order to resolve these challenges several alternative multiscale particle methods have recently been proposed. The Stochastic Particle (SP) method replaces the traditional DSMC collision step with a relaxation procedure based on Bhatnagar-Gross-Krook (BGK) type models \citep{Macrossan2001,Gallis2011,Pfeiffer2018}. A notable advantage of the SP approach is the ability to use cell sizes that are larger than the local mean free path, thereby conserving computational resources. However, the SP time step remains constrained by the local mean collision time due to the decoupling of particle motion and collisions. The Unified Stochastic Particle (USP) method has been proposed as an improvement over traditional SP approaches \citep{Fei2020,Fei2021b}. The USP scheme integrates particle motions and collisions in each time step, eliminating the aforementioned time step limitation. However, the main pitfall of these methods is that they become inaccurate in the transition and free-molecular regimes since they are both based on BGK-type models.

To circumvent this drawback hybrid particle-particle methods, such as the SP-DSMC \citep{Pfeiffer2019,Zhang2019} and USP-DSMC \citep{Fei2021a,Fei2022}, that take advantage of the computational efficiency of SP/USP in the continuum regime and the accuracy of DSMC in the rarefied regime have been proposed. In \citep{Pfeiffer2019}, Pfeiffer et al. introduced the coupled ESBGK-DSMC and ESFP-DSMC approaches in the PICLas code \citep{Ortwein2017} and evaluated their performance in a nozzle expansion case. In \citep{Fei2021a}, Fei et al. proposed a coupled USP-DSMC approach for multiscale monoatomic gas flows and later extended it to polyatomic gases in \citep{Fei2022}. The hybrid SP/USP-DSMC scheme offers several benefits over the conventional CFD-DSMC method. Firstly, a key advantage of the SP/USP-DSMC method is that the entire flow domain is handled by a single solver. This eliminates the need for a data exchange scheme between two separate solvers, thus simplifying the implementation of the SP/USP-DSMC scheme within existing DSMC solvers. Secondly, similar to the pure DSMC method, SP/USP-DSMC hybrids are inherently robust and do not suffer from the possible instabilities associated with CFD-DSMC hybrids. Thirdly, the SP/USP-DSMC approach does not require hybrid iterations for convergence in steady-state cases, resulting in greater efficiency compared to the CFD-DSMC scheme. Finally, in contrast to the CFD-DSMC approach, the SP/USP-DSMC method can employ a small number of particles per cell in time-dependent cases, enhancing its scalability. Despite the recent progress and promise of these new methods, there is currently no dedicated open-source hybrid USP-DSMC software available, representing a critical gap in the current landscape of computational tools. 

This paper presents a new solver, called uniGasFoam, developed within the OpenFOAM suite \citep{Weller1998}. Using the well-established DSMC solver dsmcFoam+ \citep{White2018} as its foundation, uniGasFoam includes hybrid SP-DSMC and USP-DSMC, as well as pure SP and USP capabilities. Furthermore, significant enhancements have been made, including advanced algorithms for transient adaptive sub-cells, adaptive non-uniform cell weighting, and adaptive global time stepping. These advances are aimed towards automatically optimising the computational cost with minimal user input, simplifying the usage of the solver and reducing potential sources of errors in the simulation results. The present work focuses on the hybrid USP-DSMC scheme, which is expected to outperform all previously proposed hybrid particle-particle schemes in terms of accuracy and efficiency. Validation of the uniGasFoam USP-DSMC module is conducted against dsmcFoam+ simulations and literature DSMC results. In addition, the computational efficiency achieved by the USP-DSMC module of uniGasFoam is reported and compared to the typical CFD-DSMC approach employed by the hybridDCFoam solver.

The rest of the paper is structured as follows: in Section~\ref{Sect:background} a brief description of the DSMC and USP methodologies is provided, while in Section~\ref{Sect:unigas}, the uniGasFoam solver and its capabilities are described in detail. Next, uniGasFoam is validated in Section~\ref{Sect:results} based on four benchmark cases: the flow past a cylinder \citep{Lofthouse2007} and over a plate \citep{Abramov2020}, the plume impingement from a conical nozzle and the time-dependent expansion into vacuum. Lastly, some concluding remarks to summarize the current work and indicate future code development are provided in Section~\ref{Sect:conclusions}.

\section{Theoretical background} \label{Sect:background}
This section provides a brief description of the DSMC, SP, and USP methods. The discussion focuses mainly on their collision schemes and distinctive attributes.

\subsection{Direct Simulation Monte Carlo} \label{Subsect:dsmc}
The DSMC method proposed by Bird \citep{Bird1994} is a stochastic particle-based method for solving the Boltzmann equation. To describe the evolution of gas flow, DSMC follows the motion and collisions of computational particles, with each computational particle representing a large number of real gas molecules $F_N$. Particle motion and collisions are decoupled over a computational time step $\Delta t$, which must be smaller than the local mean collision time to ensure accuracy in the measured macroscopic properties. The free motion part is deterministic, with each particle travelling a distance proportional to its velocity when no external forces act on it. On the other hand, gas-surface interactions and intermolecular collisions, are tackled in a stochastic manner. The most commonly used gas-surface interaction model is the diffuse-specular model \citep{Maxwell1878}, however, more advanced scattering models, such as the Cercignani-Lampis model \citep{Cercignani1971}, have been proposed in the literature. Concerning intermolecular interactions, the most widely employed collision technique is the No Time Counter (NTC) \citep{Bird1994}. In this scheme, the maximum number of binary collisions in each computational cell is calculated as

\begin{equation} \label{eq:1}
{N_C} = {1 \over 2}{{N\left( {N - 1} \right){F_N}{{\left( {{\sigma _T}{c_r}} \right)}_{\max }}\Delta t} \over {{V_C}}},
\end{equation} 

\noindent where $N$ denotes the instantaneous number of computational particles in the cell, $V_C$ is the cell volume, while the quantity ${\left( {{\sigma _T}{c_r}} \right)_{\max }}$ is the maximum value of the product of the collision cross-section $\sigma_T$ and the relative velocity $c_r$ observed in the cell throughout the simulation. Then, a total of $N_C$ particle pairs are selected randomly from the same cell and collide with probability $p_{col}$ calculated using
\begin{equation} \label{eq:2}
{p_{col}} = {{{\sigma _T}{c_r}} \over {{{\left( {{\sigma _T}{c_r}} \right)}_{\max }}}},
\end{equation} 
\noindent where ${{\sigma _T}{c_r}}$ is the product of the collision cross-section and the relative velocity of the selected particle pair. Notably, the quantity ${\left( {{\sigma _T}{c_r}} \right)_{\max }}$ only affects the number of collision trials, without affecting the actual number of collisions executed.

The implementation of the NTC scheme commonly involves the subdivision of computational cells into sub-cells \citep{Bird1994}. In this case, particle collision pairs are specifically chosen from within the same sub-cell to ensure that the collision distance remains smaller than the local mean free path. To accurately resolve intermolecular collisions several potentials have been proposed, with Hard Sphere (HS), Variable Hard Sphere (VHS), and Variable Soft Sphere (VSS) being the most widely used ones \citep{Bird1994,Koura1991}. Finally, in DSMC, the macroscopic quantities of interest such as the gas density, temperature, and velocity are calculated by sampling the molecular properties of the computational particles.

\subsection{Stochastic Particle and Unified Stochastic Particle} \label{Subsect:usp}
The SP and USP methods are probabilistic particle-based schemes for solving BGK-type kinetic model equations. Kinetic models approximate the Boltzmann equation by replacing the exact collision operator with simplified expressions \citep{Bhatnagar1954,Holway1966,Shakhov1972}. The SP and USP algorithms are similar to the DSMC one. However, the SP and USP methods replace the DSMC collision step with a relaxation process, where after the free-motion step, particles relax to a target equilibrium distribution $f_{eq}$ with some relaxation probability $p$. The advantage of the USP method is that it uniquely integrates molecular motions and collision effects into a single computational time step, enabling its application with significantly larger time steps compared to the previously proposed SP method. The SP and USP target distribution and relaxation probability depend on the kinetic model. Without the loss of generality, the relaxation probability $p_{rel}$ for the SBGK \citep{Shakhov1972} model is

\begin{equation} \label{eq:3}
p_{rel} = 1 - {e^{ - v\Delta t}}, \quad v={P \over \mu },
\end{equation} 

\noindent where $v$ is the collision frequency, with $\Pr$, $P$ and $\mu$ denoting the gas Prandtl number, pressure, and viscosity respectively. The SBGK target distributions for the SP and USP methods are written as

\begin{equation} \label{eq:4}
f_{eq}^{SP} = {f_M}\left[ {1 + \left( {1 - \Pr } \right){{{q_k}{C_k}} \over {PRT}}\left( {{{{C^2}} \over {5RT}} - 1} \right)} \right],
\end{equation} 

\begin{equation} \label{eq:5}
f_{eq}^{USP} = {f_M}\left\{ {1 + A{{{\sigma _{ij}}{C_{ < i}}{C_{j > }}} \over {2PRT}} + B{{{q_k}{C_k}} \over {PRT}}\left[ {{{{C^2}} \over {5RT}} - 1} \right]} \right\},
\end{equation} 

\noindent where $f_M$ is the Maxwellian distribution, $R$ is the specific gas constant, $T$ is the gas temperature, $\bf{C}=\bf{c}-\bf{u}$ is the peculiar velocity vector, with $\bf{c}$ and $\bf{u}$ denoting the molecular and macroscopic velocity vectors, while $\bf{q}$ and $\overline{\overline \sigma }$ denote the heat flux vector and stress tensor, respectively. The constants $A$ and $B$ for the USP scheme are given by

\begin{equation} \label{eq:6} 
A = 1 - {{v\Delta t} \over 2}\coth \left( {{{v\Delta t} \over 2}} \right),\quad B = 1 - {{\Pr v\Delta t} \over 2}\coth \left( {{{v\Delta t} \over 2}} \right).
\end{equation} 

Note that, since the SP-SBGK and USP-SBGK target distributions are polynomial with respect to the molecular velocity, they can become negative. Although the negativity of the distribution function is not normally significant, it can become important in the case of large heat fluxes and/or stresses. This problem is compounded by the inherent statistical nature of the proposed methodology, where instantaneous samples of heat fluxes and stresses can become abnormally large. To circumvent this, an exponential time-averaging scheme \citep{Feng2023} is employed

\begin{equation} \label{eq:7}
{\varphi ^t} = \theta {\varphi ^t} + \left( {1 - \theta } \right){\varphi ^{t - 1}},\quad \varphi  = {q_k},{\sigma _{ij}},
\end{equation} 

\noindent where ${\varphi ^t}$ and $\varphi ^{t - 1}$ denote the averaged macroscopic quantity in the current and previous time step, respectively, while $\theta$ is the time-averaging coefficient.

To ensure momentum and energy conservation all particle velocities are rescaled after the relaxation step according to

\begin{equation} \label{eq:8}
c_k^r = {u_k} + \left( {c_k^p - u_k^p} \right)\sqrt {{T \over {{T^p}}}},
\end{equation} 

\noindent where $c_k^r$ is the rescaled molecular velocity, $u_k$ and $T$, are the pre-relaxation macroscopic velocity and temperature, while $c_k$, $u_k^p$ and $T^p$ are the post-relaxation molecular velocity, and macroscopic velocity and temperature, respectively.

In the SP and USP methods, particle relaxation depends solely on the local target distribution function, defined by the local macroscopic quantities. In the continuum regime, changes in the macroscopic quantities only occur over distances larger than the local mean free path. Thus, the SP and USP methods permit the use of computational cells larger than the local mean free path in the continuum regime. However, to achieve second-order accuracy, the macroscopic quantities and thus the underlying target distribution function must be reconstructed during the relaxation process at the location of the simulated particle, which can be achieved by a linear or random interpolation \citep{Fei2021b}. Additionally, previous works have shown that the USP method has second-order temporal accuracy and can employ a computational time step larger than the local mean collision time \citep{Fei2020}. These attributes position the SP and USP methods as highly suitable for coupling with DSMC.


\section{The uniGasFoam solver} \label{Sect:unigas}

\subsection{Downloading and installing uniGasFoam} \label{Sect:install}
The developed solver can be downloaded from the associated CPC library entry, as well as from the provided GitHub repository, where future additions and advancements will be added. It is assumed that users will already have a suitable version of OpenFOAM installed. More specifically, the CPC library entry is compatible with OpenFOAM-v2406,  while the GitHub repository will be kept up to date with the latest OpenFOAM version. Instructions for installing and building uniGasFoam can be found in \textit{doc/\-uni\-Gas\-Foam\-INSTALL.md} within the main repository directory.

The source files for the libraries are located in \textit{src/\-lagrangian} and the executable file for running the solver can be found in \textit{application/\-solvers/\-discrete\-Methods/\-uniGasFoam}. In addition, all the benchmark cases presented in this work can be found in the \textit{tutorials/\-uniGasFoam} directory.

\subsection{Using uniGasFoam} \label{Sect:usage}
Using uniGasFoam will be relatively straightforward for users who are already familiar with OpenFOAM solvers. Since the presented solver has been developed based on dsmcFoam+ \citep{White2018}, the particle initialization, time control, mesh generation, and post-processing procedures, are very similar and can be found in Section 4 of \citep{White2018}. Furthermore, a detailed description of case file structure and dictionary entries is provided in \textit{doc/\-uni\-Gas\-Foam\-GUIDE.md} as a guide to the usage of uniGasFoam. Hence, in the following sections, only the main changes and advancements of uniGasFoam compared to dsmcFoam+ will be presented.

\subsection{Collision models} \label{Subsect:collisions}
The uniGasFoam solver is able to perform hybrid SP-DSMC and USP-DSMC, as well as pure DSMC, SP and USP gas-flow simulations. When the pure DSMC model is implemented the solver is identical to dsmcFoam+, except for the advancements detailed in Section~\ref{Subsect:adapt}, and can handle monoatomic and polyatomic gases, as well as gas mixtures and chemical reactions. However, when the rest of the models are employed, uniGasFoam is currently limited to single monoatomic non-reacting gases. The implemented collision model and its associated properties are provided in the \textit{[case]/\-constant/\-uniGasProperties} file. An example of the collision model definition is given here: 

\begin{scriptsize}
	\begin{verbatim}
    1  collisionModel              hybrid;
    2  relaxationCollisionModel    unifiedStochastiParticleSBGK;
    3  binaryCollisionPartnerModel noTimeCounter;
    4  binaryCollisionModel        variableHardSphere;
    5  collisionProperties
    6  {
    7      Tref                    273;
    8      macroInterpolation      true;
    9      theta                   1e-3;
    10 }
	\end{verbatim}
\end{scriptsize}

Referring to the example above, Line 1 defines the collision model and can take three different values, namely \textit{binary} for pure DSMC, \textit{relaxation} for pure SP/USP and \textit{hybrid} for a coupled SP/USP-DSMC simulation. In Line 2 the SP or USP relaxation model is defined, while in Lines 3 and 4, the DSMC collision partner selection scheme and intermolecular potential are provided. The collision properties are given in Lines 6-10. More specifically, the particle diameter reference temperature is provided in Line 7, the SP/USP spatial interpolation is enabled/disabled in Line 8, and the time-averaging coefficient is defined in Line 9.

\subsection{Domain decomposition} \label{Subsect:decomposition}
In the case that uniGasFoam is used for hybrid SP/USP-DSMC simulations, the domain decomposition module is activated. During the decomposition process, the mesh cells are split into, either continuum (SP/USP) or rarefied (DSMC) cells. Based on the cell type, particles are then treated differently in terms of the collision model. Specifically, during the collision step, stochastic relaxations are performed in the continuum (SP/USP) cells, while stochastic binary collisions are performed in the rarefied (DSMC) cells. The domain decomposition method and the associated numerical properties are defined within the \textit{[case]/\-system/\-hybrid\-Decomposition\-Dict} dictionary. An example of this dictionary is provided here: 

\begin{scriptsize}
	\begin{verbatim}
    1  decompositionModel          localKnudsen;
    2 
    3  timeProperties
    4  {
    5      decompositionInterval           20;
    6      resetAtDecomposition            true;
    7      resetAtDecompositionUntilTime   2e-3;
    8  }
    9
    10 localKnudsenProperties
    11 {
    12     breakdownMax        0.05;
    13     theta               0.2;
    14     smoothingPasses     5;
    15 }
	\end{verbatim}
\end{scriptsize}

In the above example, the domain decomposition model is defined in Line 1, while in Lines 3-8 and 10-15 the time and decomposition properties are provided, respectively. Currently, only the \textit{localKnudsen} model defined in Line 1 can be used, which corresponds to the gradient local length Knudsen number breakdown criterion \citep{Wang2003A}, written as:
\begin{equation} \label{eq:9}
\mathrm{Kn}_{GLL} = \lambda \max \left[ {{{\left\| {\nabla \rho } \right\|} \over \rho },{{\left\| {\nabla T} \right\|} \over T},{{\left\| {\nabla u} \right\|} \over {\max \left( {u,{v_0}} \right)}}} \right].
\end{equation}
Here, $\lambda$ is the local gas mean free path, $\rho$, $T$, and $u$ denote the local macroscopic density, temperature, and velocity magnitude, while $v_0=\sqrt{2RT}$ is the most probable molecular speed.
It is noted that additional breakdown criteria, such as the simplified Chapman-Enskog parameter \citep{Garcia1998}, may be easily incorporated in the future. The domain is decomposed every 100 time steps by default, but can be overridden by the user to account for extreme flow-state changes as shown in Line 5. The averaging of the macroscopic quantities used in the domain decomposition module, such as the gas density, temperature and velocity, can be enabled or disabled as the simulation runs in Line 6. Additionally, by setting \textit{reset\-At\-Decomposition} to \textit{true} and specifying the optional \textit{reset\-At\-Decomposition\-Until\-Time} entry at Line 7, the user can determine the time point after which the averaging of macroscopic quantities starts. The continuum breakdown threshold is provided in Line 12, with a typical value of 0.05 being proposed in the literature \citep{Wang2003A}. Cells with $\mathrm{Kn}_{GLL}>0.05$ are classified as rarefied (DSMC) cells and otherwise as continuum cells (SP/USP). To have a smooth interface between the continuum and rarefied regions, exponential time-averaging is implemented for measuring ${\mathrm{Kn}_{GLL}}$ with the time-averaging coefficient defined in Line 13. In addition, to have a smoother interface between the continuum and rarefied regions, Laplacian smoothing is employed for the density, temperature and velocity fields used to compute ${\mathrm{Kn}_{GLL}}$, with the number of smoothing passes provided in Line 14.

\subsection{Adaptive schemes} \label{Subsect:adapt}
Several adaptive schemes have been added to the uniGasFoam solver to increase its computational efficiency, automate parts of the simulation process, and eliminate user input errors. More specifically, uniGasFoam can utilise transient adaptive sub-cells, non-uniform particle cell weighting, as well as adaptive global time stepping. Situations where the implemented adaptive schemes are required include flows with widely varying density or velocity across the flow domain, as well as transient flows where the sub-cell discretisation and time step need to evolve throughout the simulation for optimum computational efficiency and accuracy.

\begin{figure}[]
\centerline{
\includegraphics*[width=0.7\textwidth, keepaspectratio=true]{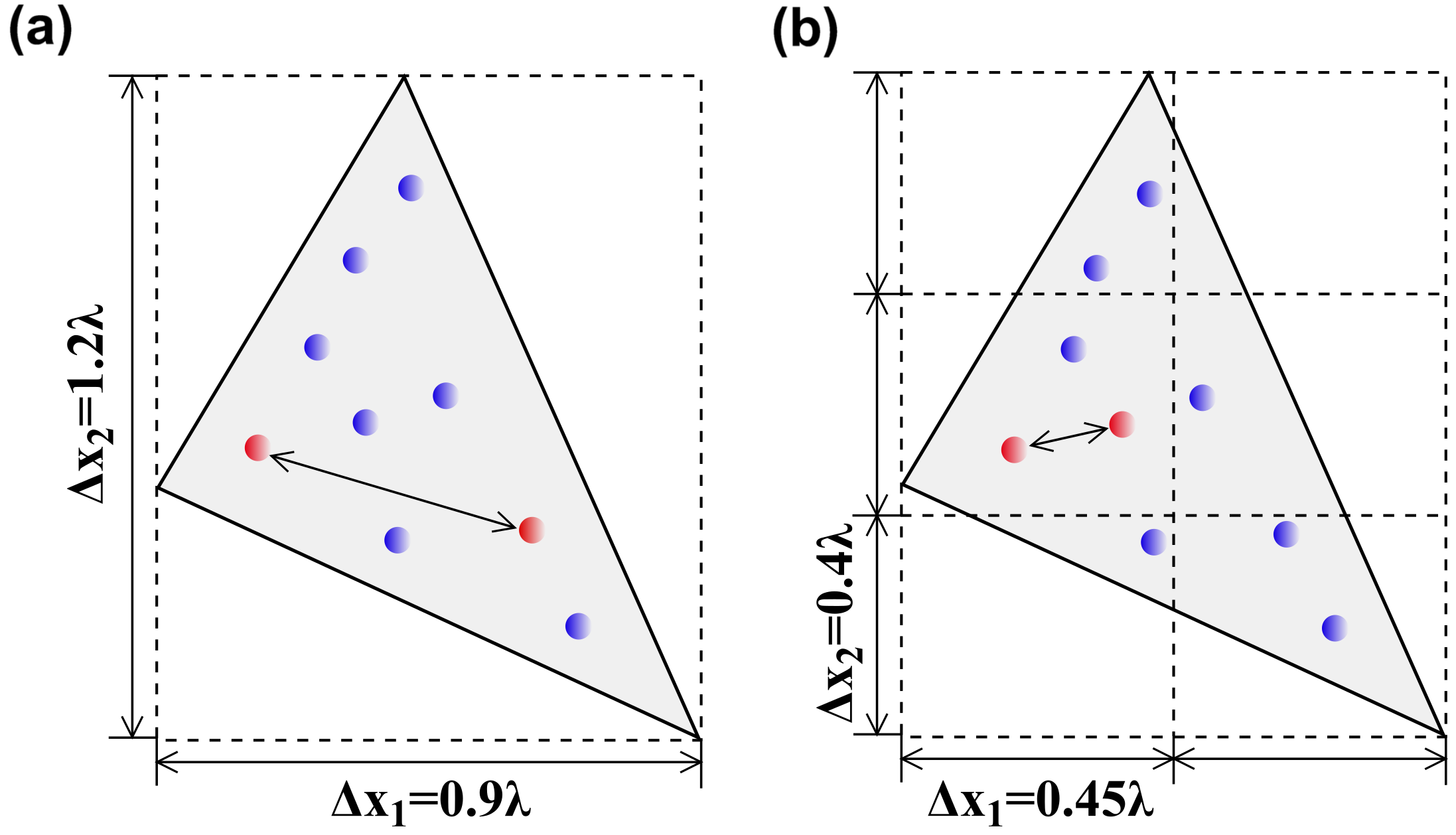}
}
\caption{Schematic showing the division of an unstructured cell into Cartesian sub-cells using the TAS algorithm. In (a), no sub-cells are implemented, and any particle pair within the cell can be chosen for collision. In (b), the cell is divided into Cartesian sub-cells, and only particle pairs within the same sub-cell can be chosen for collision.}
\label{fig:TAS}
\end{figure}

The transient adaptive sub-cells (TAS) algorithm eliminates user-input error in the mesh creation process by splitting cells larger than the local mean free path into Cartesian sub-cells as shown in Fig.~\ref{fig:TAS}. Cell splitting is performed every 20 time steps by default, and can be overridden by the user to account for extreme flow state changes. The created sub-cells are used only for binary collision pair selection, ensuring that the collision distance between two particles is smaller than a user-defined fraction of the mean free path. The number of sub-cells in each Cartesian direction $N_{SC,i}$ is calculated as
\begin{equation} \label{eq:10}
{N_{SC,i}} = \left\lfloor {{{\Delta {x_i}} \over {{f_s}\lambda }}} \right\rfloor  + 1,
\end{equation}
\noindent where $\Delta {x_i}$ denotes the cell size in each direction, $\lambda$ is the local mean free path and $f_s$ is the desired sub-cell size to mean free path ratio. The TAS algorithm is implemented only for DSMC cells, as the SP and USP cell size is not constrained by the local mean free path. It is noted that cells that are not further subdivided into sub-cells are considered to contain a single sub-cell, which coincides with the original cell.

The non-uniform cell weighting algorithm is employed to keep a constant number of particles per sub-cell, thus decreasing computational cost while maintaining accuracy. To achieve this, a different weight $W_C$ is assigned to each cell
\begin{equation} \label{eq:11}
{W_C} = {{nV} \over {{N_{PPSC}}{N_{SC}}}},
\end{equation}
\noindent where $n$ and $V$ are the cell number density and volume, $N_{PPSC}$ denotes the desired number of particles per sub-cell, while $N_{SC}$ is the total number of sub-cells contained in the cell. As a particle moves from an initial cell with weight ${\rm{W}}_C^{t - 1}$ to a final cell with weight ${\rm{W}}_C^{t}$ it is cloned ${{{\rm{W}}_C^{t - 1}} \mathord{\left/
 {\vphantom {{{\rm{W}}_C^{t - 1}} {{\rm{W}}_C^t}}} \right.
 \kern-\nulldelimiterspace} {{\rm{W}}_C^t}} - 1$ times when ${\rm{W}}_C^t \le {\rm{W}}_C^{t - 1}$, or is deleted with probability ${{{\rm{W}}_C^{t - 1}} \mathord{\left/
 {\vphantom {{{\rm{W}}_C^{t - 1}} {{\rm{W}}_C^t}}} \right.
 \kern-\nulldelimiterspace} {{\rm{W}}_C^t}}$ otherwise. The presented scheme can be applied once at the start of the simulation based on the initial number density or employed adaptively throughout the simulation, using the number density averaged at a user-defined interval, which is set to 20 time steps by default.

The adaptive global time stepping algorithm eliminates user-input error by adaptively implementing the correct time step throughout the simulation. The optimal global time step is recalculated every 20 time steps, by default, and can be overridden by the user if necessary to account for extreme flow-state changes. In SP and DSMC cells both the mean collision and cell traversal time must be resolved, and thus the optimal time step is calculated as
\begin{equation} \label{eq:12}
\Delta {t_{\mathrm{DSMC}}} = \Delta {t_{\mathrm{SP}}} = \min \left[ {{f_t}\tau ,\mathrm{Co}{{\Delta {x_i}} \over {\max \left( {{u_i},{v_0}} \right)}}} \right],
\end{equation}
\noindent where $\tau$ is the local mean collision time, $\Delta {x_i}$ is the cell size, and $u_i$ are the macroscopic velocity components, respectively, while $f_t$ and $\mathrm{Co}$ denote the user-defined time step to mean-collision-time ratio and Courant number, respectively. In USP cells, only the mean cell-traversal time needs to be resolved, hence the optimal time step is computed by 
\begin{equation} \label{eq:13}
\Delta {t_{\mathrm{USP}}} = \min \left[ {\mathrm{Co}{{\Delta {x_i}} \over {\max \left( {{u_i},{v_0}} \right)}}} \right].
\end{equation}

In the case of a hybrid USP-DSMC simulation, the global time step is calculated as the minimum of the USP and DSMC time steps
\begin{equation} \label{eq:14}
\Delta t = \min \left( {\Delta {t_{\mathrm{DSMC}}},\Delta {t_{\mathrm{USP}}}} \right).
\end{equation}

The implemented adaptive schemes and their associated properties are provided in the \textit{[case]/\-constant/\-uniGasProperties} file. An example of the adaptive scheme definition is presented here: 

\begin{scriptsize}
	\begin{verbatim}
    1  // Cell Weighting Properties
    2  // ~~~~~~~~~~~~~~~~~~
    3  cellWeightedSimulation      	true;
    4  cellWeightedProperties
    5  {
    6      minParticlesPerSubCell   10;
    7      particlesPerSubCell      20;
    8  }
    9
    10 // Dynamic Adaptation
    11 // ~~~~~~~~~~~~~~~~~~
    12 adaptiveSimulation          true;
    13 adaptiveProperties
    14 {
    15     cellWeightAdaptation    true;
    16     subCellAdaptation       true;
    17     timeStepAdaptation      true;
    18     adaptationInterval      20;
    19     maxSubCellSizeMFPRatio  0.5;
    20     maxTimeStepMCTRatio     0.2;
    21     maxCourantNumber        0.5; 
    22 }
	\end{verbatim}
\end{scriptsize}

Referring to the above example, in Lines 3-8 the non-uniform cell weighting algorithm properties are given. The cell-weighting algorithm is enabled or disabled in Line 3, while in Lines 6 and 7 the minimum and desired number of particles per sub-cell are defined. In Lines 12-21 the adaptive algorithm properties are provided. The adaptive schemes are enabled or disabled in Lines 12, 15, 16 and 17. The adaptive quantities, namely the number of sub-cells, the cell weights and the global time step are recalculated by default every 20 time steps but can be overridden by the user to account for extreme flow state changes as shown in Line 18. In Line 19, the maximum allowed sub-cell size to local mean free path ratio for the TAS algorithm is provided. The maximum allowed time step to local mean free path ratio and the maximum Courant number are defined in Lines 20 and 21, respectively.  

\subsection{Algorithm overview} \label{Subsect:algorithm}

\begin{figure}[]
\centerline{
\includegraphics*[width=0.7\textwidth, keepaspectratio=true]{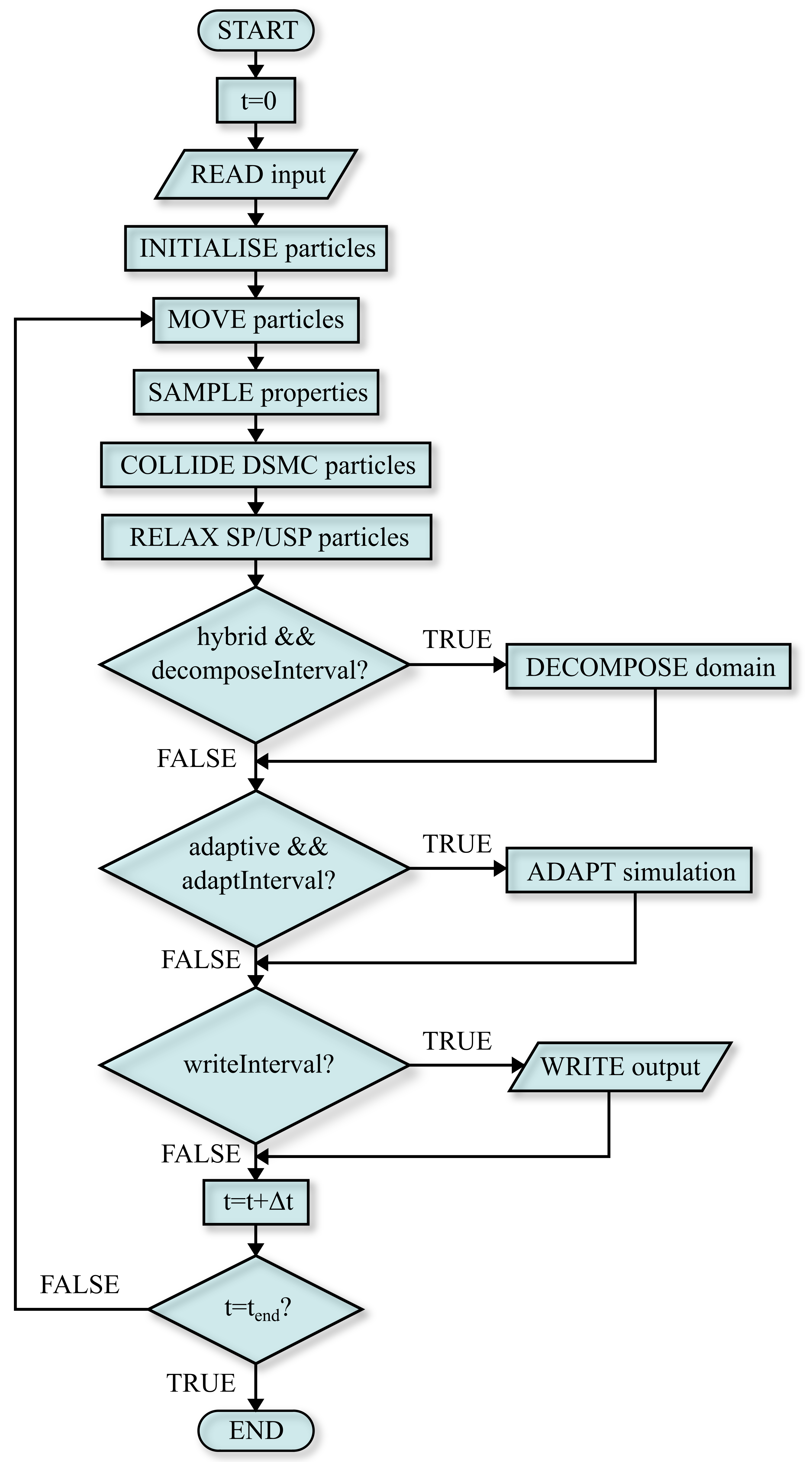}
}
\caption{Flow chart of uniGasFoam solver.}
\label{fig:flowchart}
\end{figure}

Here, a brief overview of the uniGasFoam solver is presented, with the algorithm flow chart shown in Fig.~\ref{fig:flowchart}. Initially, simulation input data are read from their respective dictionaries, and computational particles are stochastically initialised within the flow domain. Upon starting the simulation, each time step follows a structured sequence of steps. First, particles are moved deterministically, and boundary conditions are applied. Next, macroscopic quantities of interest such as density, temperature, and velocity are sampled. In DSMC cells, binary collisions are performed, while in SP and USP cells, particle relaxations to the underlying target distribution are executed. In the case of a hybrid simulation, the flow domain is decomposed into continuum and rarefied regions based on the applied breakdown criteria, at a user-defined frequency. Similarly, the number of sub-cells, cell weights, and global time step are recalculated for adaptive simulations at a user-specified interval. Subsequently, the macroscopic quantities of interest are averaged and outputted. The algorithm then proceeds to the next time step and repeats the described process until the final simulation time is reached.

\section{Results and discussion} \label{Sect:results}
In this section, the results obtained by the hybrid USP-DSMC module of the uniGasFoam solver are presented and discussed. These results are validated based on pure DSMC simulation results generated by the dsmcFoam+ \citep{White2018} solver and benchmark data reported in the literature. A comparison is also made our CFD-DSMC hybrid solver hybridDCFoam \citep{Vasileiadis2024}, developed within the OpenFOAM framework, and the computational savings are estimated. Four benchmark cases have been chosen, namely, the flow past a cylinder \citep{Lofthouse2007} in Section~\ref{Subsect:cylinder}, flow over a plate \citep{Abramov2020} in Section~\ref{Subsect:plate}, nozzle plume impingement in Section~\ref{Subsect:nozzle}, and transient expansion into vacuum in Section~\ref{Subsect:expansion}. Hence, the accuracy of the developed hybrid solver is validated in a wide range of applications.

All approaches used the same adaptive schemes with identical numerical parameters in order to provide a fair comparison. More specifically, the maximum allowed sub-cell size was equal to 1/2 of the local mean free path, while the adaptive global time stepping algorithm was employed with a maximum time step to mean collision time ratio equal to 1/5 and a maximum CFL number of 1/2. The desired number of particles per sub-cell was set to 20 in all cases, unless otherwise specified. In addition, the domain decomposition for the uniGasFoam and hybridDCFoam solvers was always performed based on a gradient local length Knudsen threshold value of 0.05. Finally, all presented simulations were performed on an AMD EPYC 7742 @2.25GHz node on the UK National Supercomputing Service ARCHER2. 

\subsection{Flow past cylinder} \label{Subsect:cylinder}

\begin{figure}[]
\centerline{
\includegraphics*[width=0.6\textwidth, keepaspectratio=true]{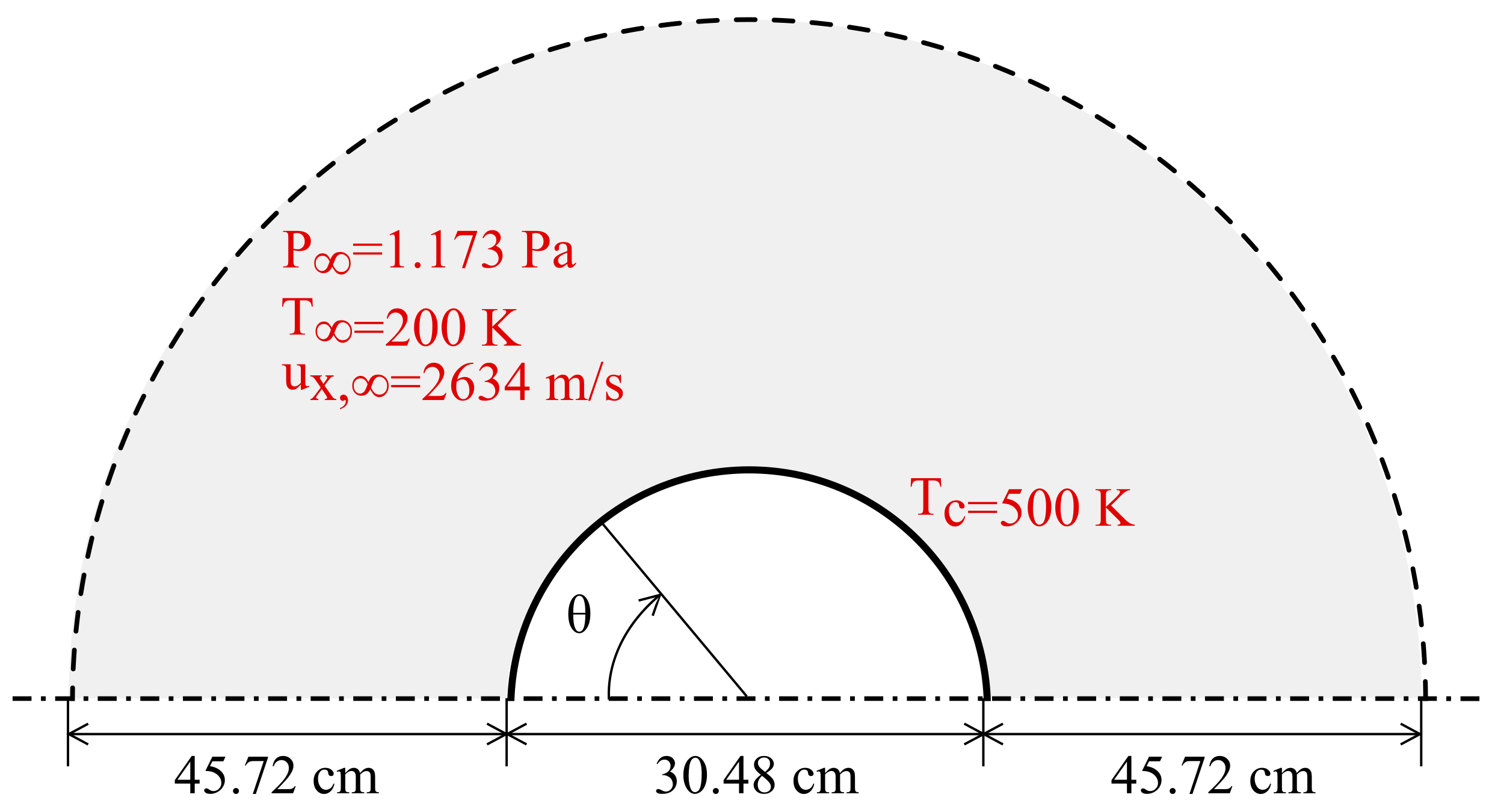}
}
\caption{Configuration of the hypersonic flow past a cylinder benchmark case.}
\label{fig:cylinderConfig}
\end{figure}

The first benchmark case is the hypersonic flow over a cylinder. In Fig.~\ref{fig:cylinderConfig}, the flow configuration and the applied boundary conditions are presented. Taking advantage of the flow symmetry, only half of the domain is simulated to reduce the required computational cost. The simulated gas is argon and the NTC collision scheme with the VHS potential is implemented. The gas properties are taken in accordance to \citep{Lofthouse2007}. The cylinder diameter $D$ is equal to 0.3048 m, while the distance between the cylinder and the freestream boundaries $S$ was set to 1.5$D$. The freestream argon pressure $P_{\infty}$, temperature $T_{\infty}$, and longitudinal velocity $u_{x,\infty}$ are 1.173 Pa, 200 K, and 2634 m/s. These freestream conditions correspond to a Knudsen number of 0.01 based on the cylinder diameter and a Mach number of 10. The cylinder surface is modelled as purely diffuse with a constant temperature of $T_c$=500 K. The flow domain is discretized by employing a structured mesh of 100$\times$200 non-uniform hexahedral cells. Steady-state sampling for the macroscopic quantities commenced at 1 ms, and a total of 10 ms were simulated to reduce the statistical noise.

\begin{figure}[]
\centerline{
\includegraphics*[width=0.6\textwidth, keepaspectratio=true]{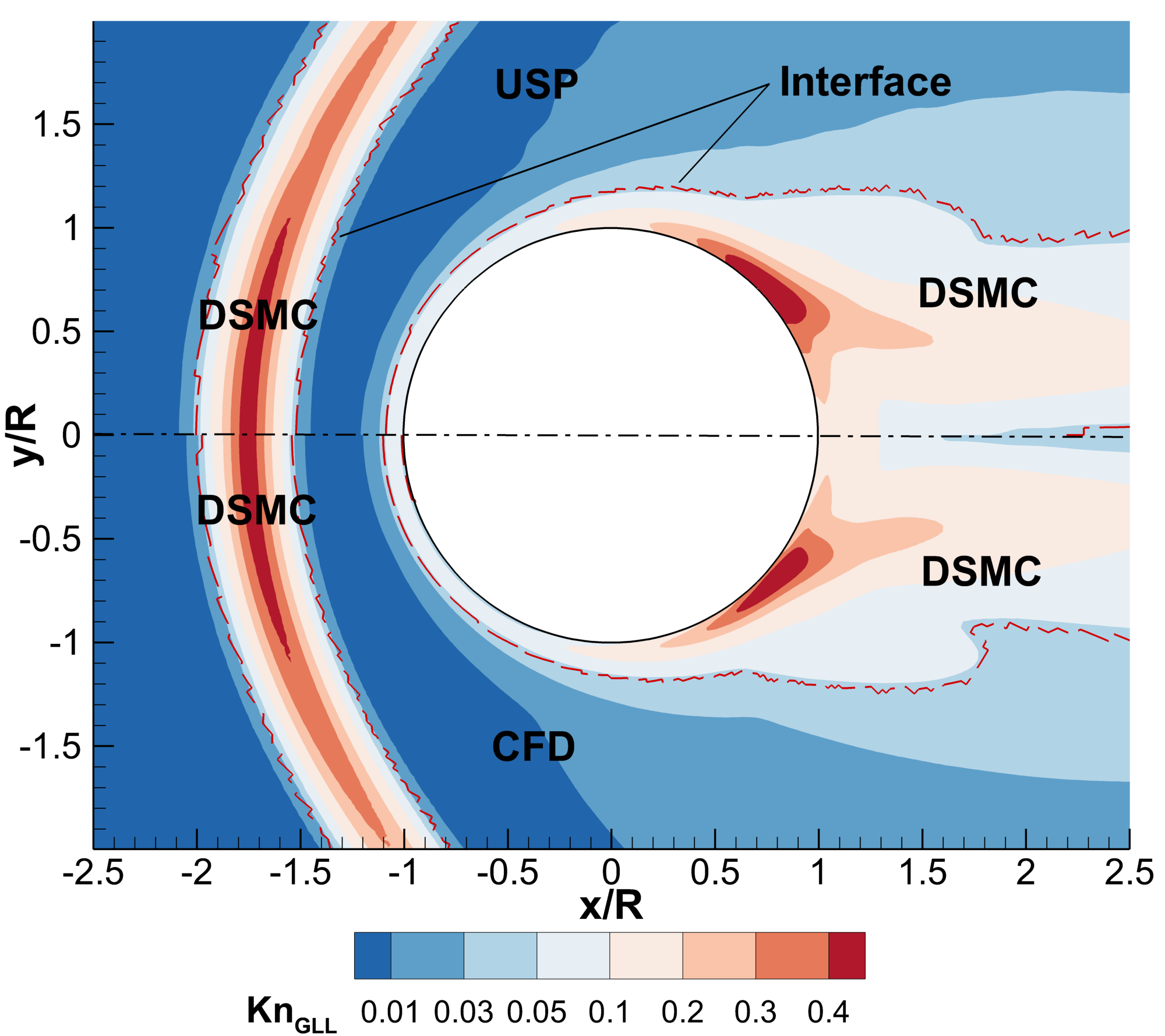}
}
\caption{Maximum gradient local length Knudsen number $\mathrm{Kn}_{GLL}$, along with continuum-rarefied interface predicted by uniGasFoam (top) and hybridDCFoam (bottom), for the flow past a cylinder case.}
\label{fig:cylinderSliceKn}
\end{figure}

In order to visualise the continuum and rarefied regions, as well as to confirm that uniGasFoam and hybridDCFoam predict a similar domain decomposition, the maximum gradient local length Knudsen number $\mathrm{Kn}_{GLL}$ calculated by uniGasFoam and hybridDCFoam, along with the continuum-rarefied interfaces are presented in Fig.~\ref{fig:cylinderSliceKn}. As expected, the predicted continuum-rarefied interfaces closely track the contour $\mathrm{Kn}_{GLL}$=0.05. The rarefied region captures the bow shock, the Knudsen layer around the cylinder, and the wake area. In this particular case, the gradient local length Knudsen numbers based on the density and temperature gradients become more dominant in the bow shock and Knudsen layer regions, while the velocity gradient local length Knudsen number becomes more dominant in the wake region.

\begin{figure}[]
\centerline{
\includegraphics*[width=0.45\textwidth, keepaspectratio=true]{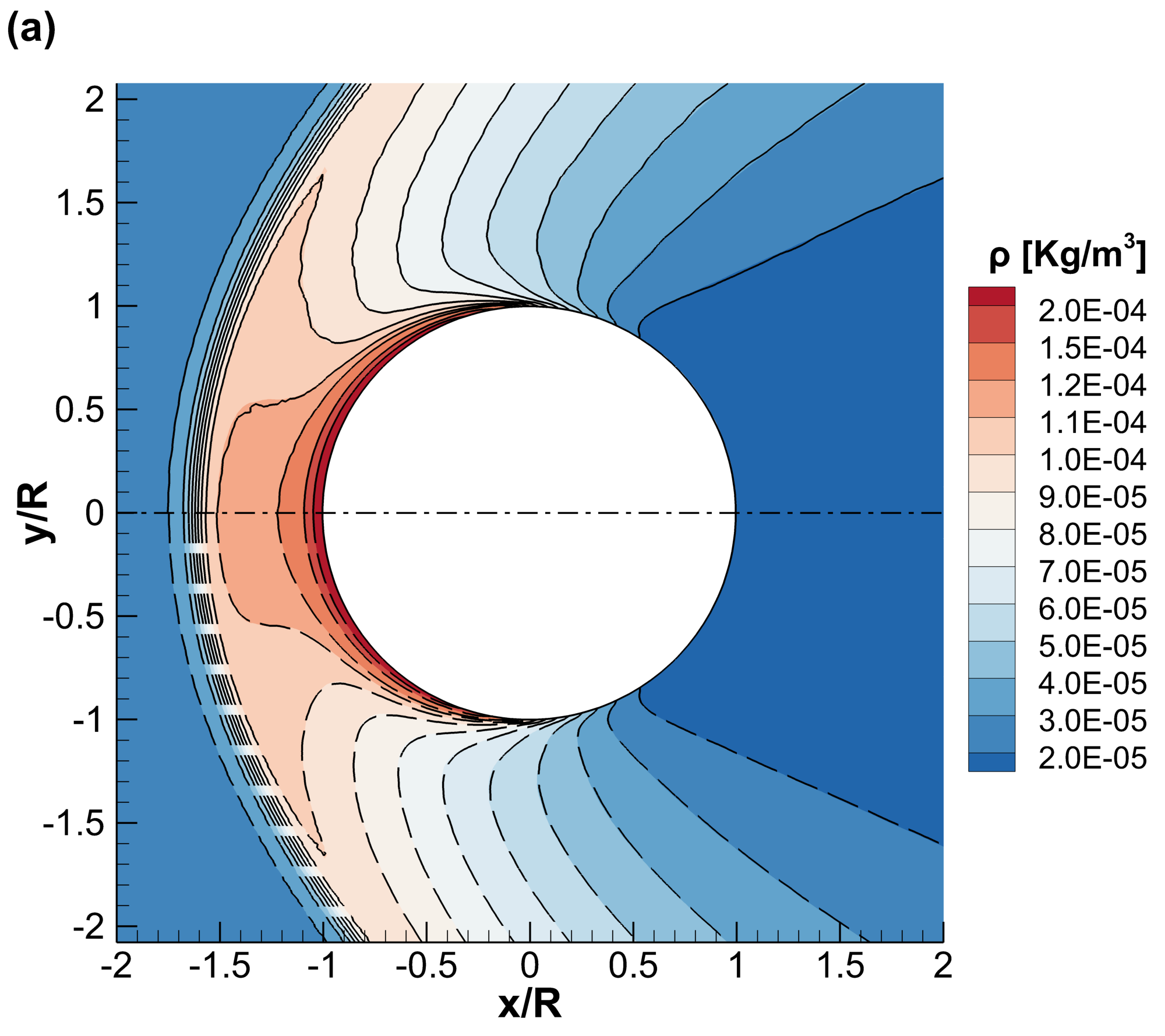}
}
\centerline{
\includegraphics*[width=0.45\textwidth, keepaspectratio=true]{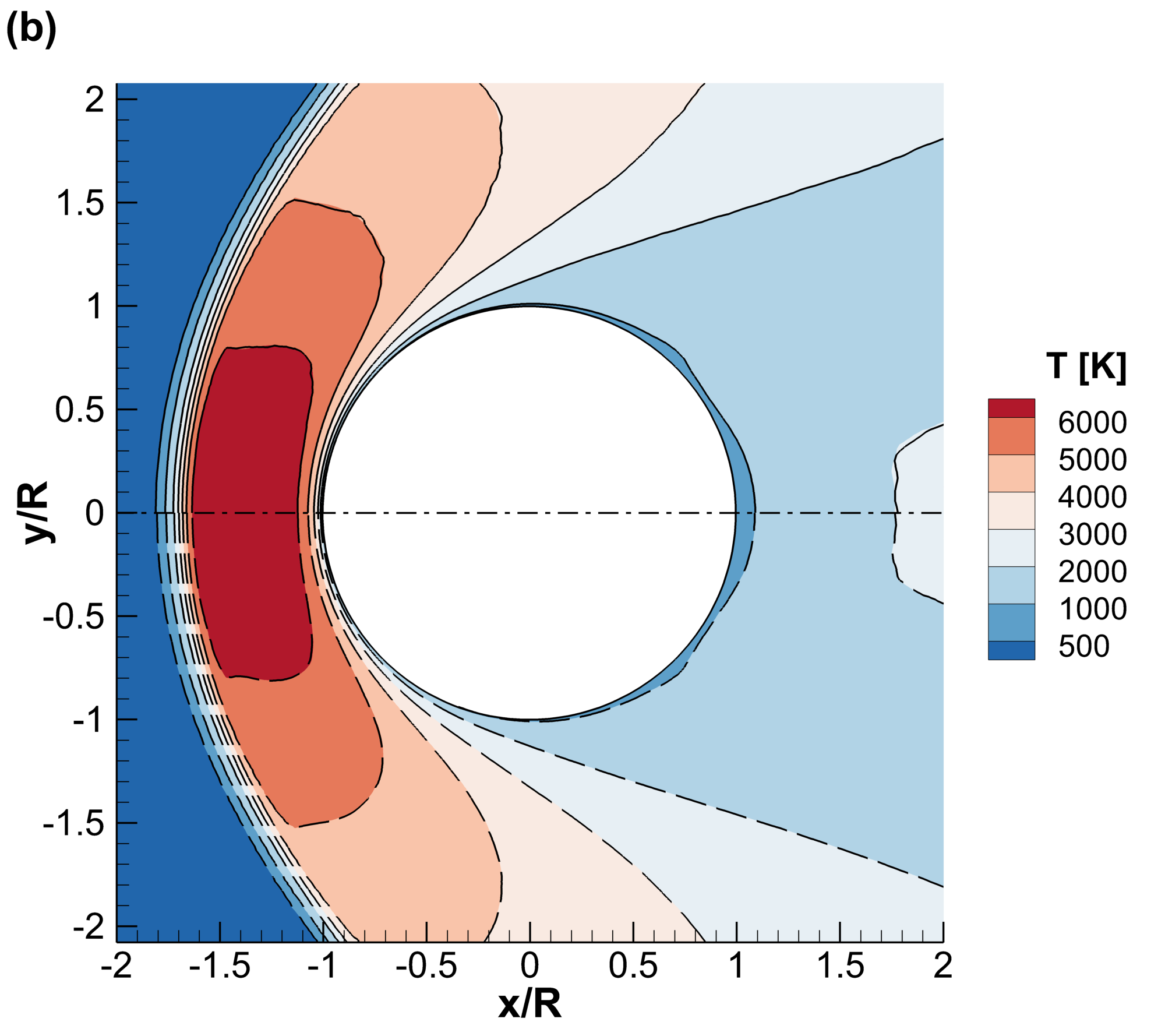}
}
\centerline{
\includegraphics*[width=0.45\textwidth, keepaspectratio=true]{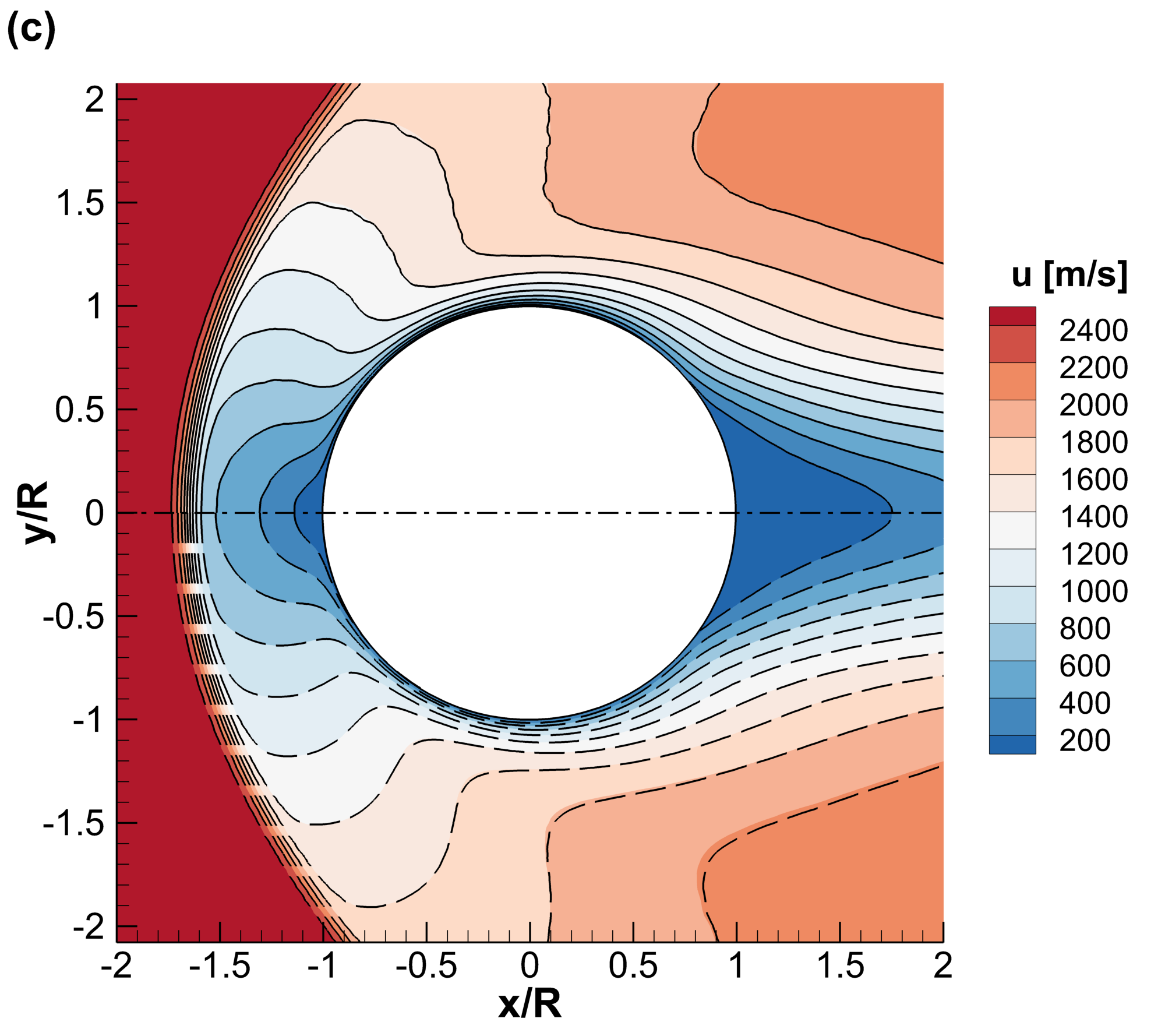}
}
\caption{Comparison of (a) density, (b) temperature, and (c) velocity magnitude obtained by uniGasFoam (solid lines), dsmcFoam+ (colour flood) and hybridDCFoam (dashed lines), for the flow past a cylinder case.}
\label{fig:cylinderSlice}
\end{figure}

Contour plots of the density, temperature, and velocity magnitude fields predicted by uniGasFoam, dsmcFoam+, and hybridDCFoam are shown in Fig.~\ref{fig:cylinderSlice}. An excellent agreement is observed between the presented methods. It is evident that the uniGasFoam solver captures the bow shock, Knudsen layer, and wake structures, thus validating that it can accurately predict the regions where non-equilibrium phenomena occur.

\begin{figure}[]
\centerline{
\includegraphics*[width=0.45\textwidth, keepaspectratio=true]{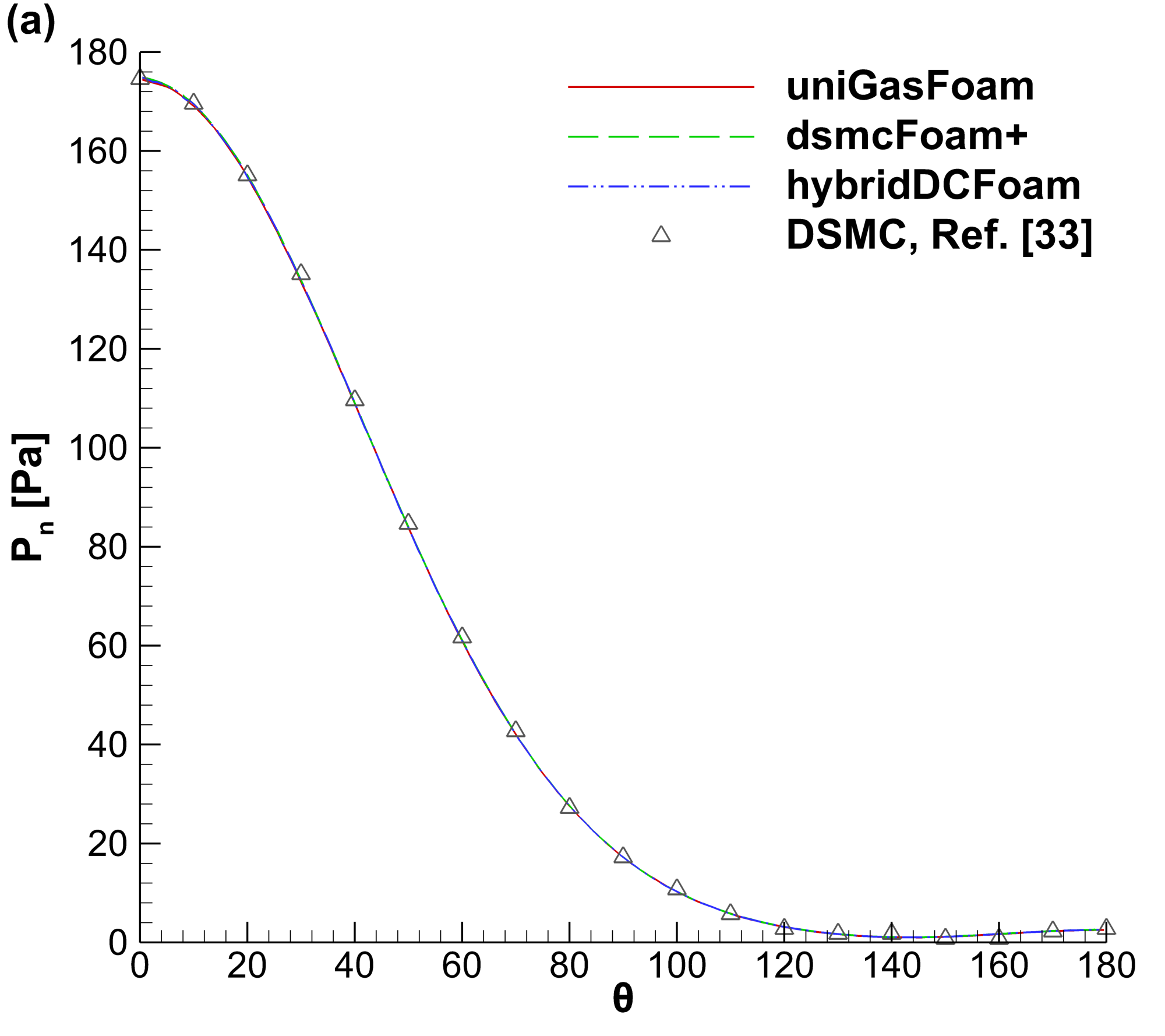}
}
\centerline{
\includegraphics*[width=0.45\textwidth, keepaspectratio=true]{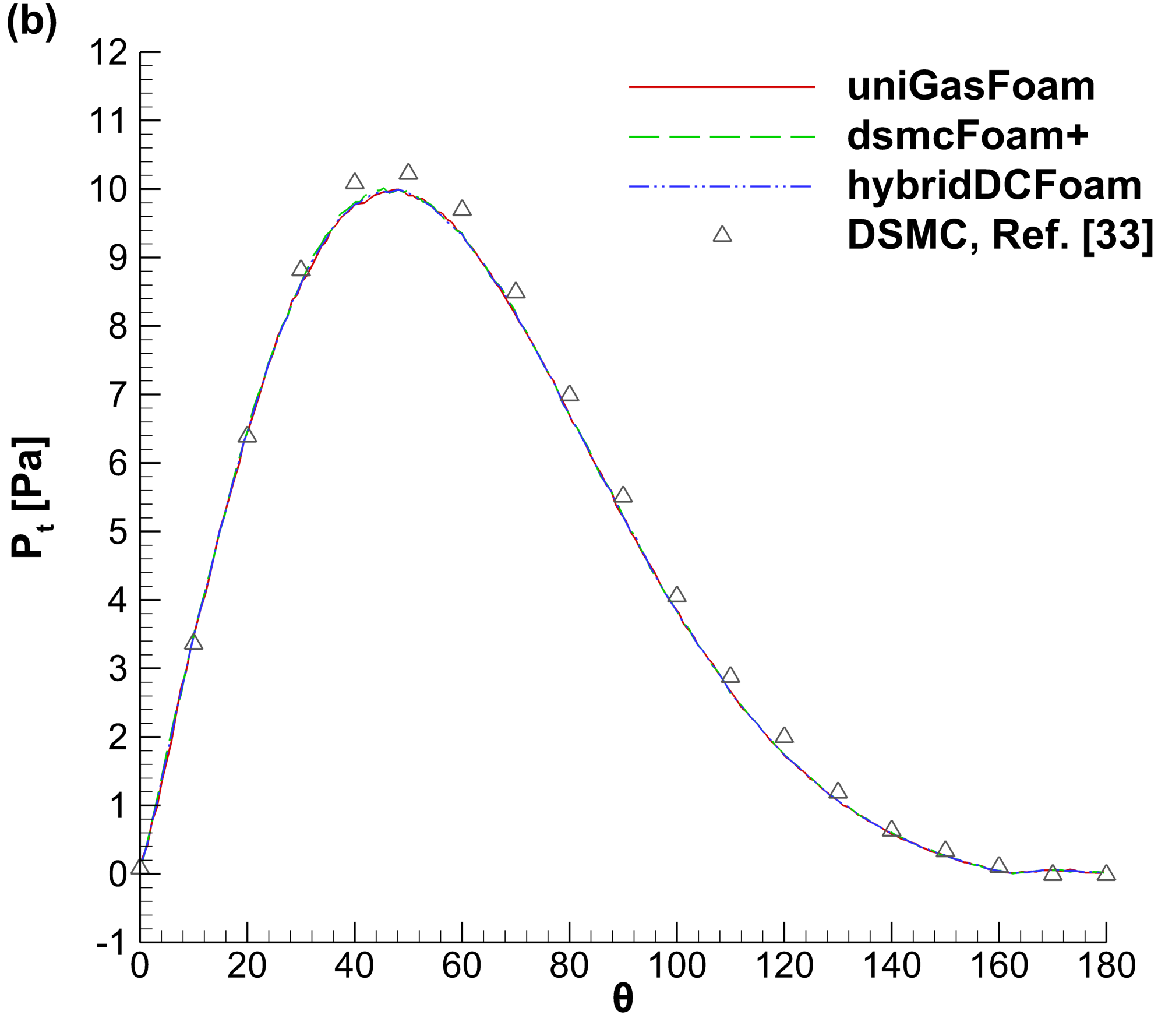}
}
\centerline{
\includegraphics*[width=0.45\textwidth, keepaspectratio=true]{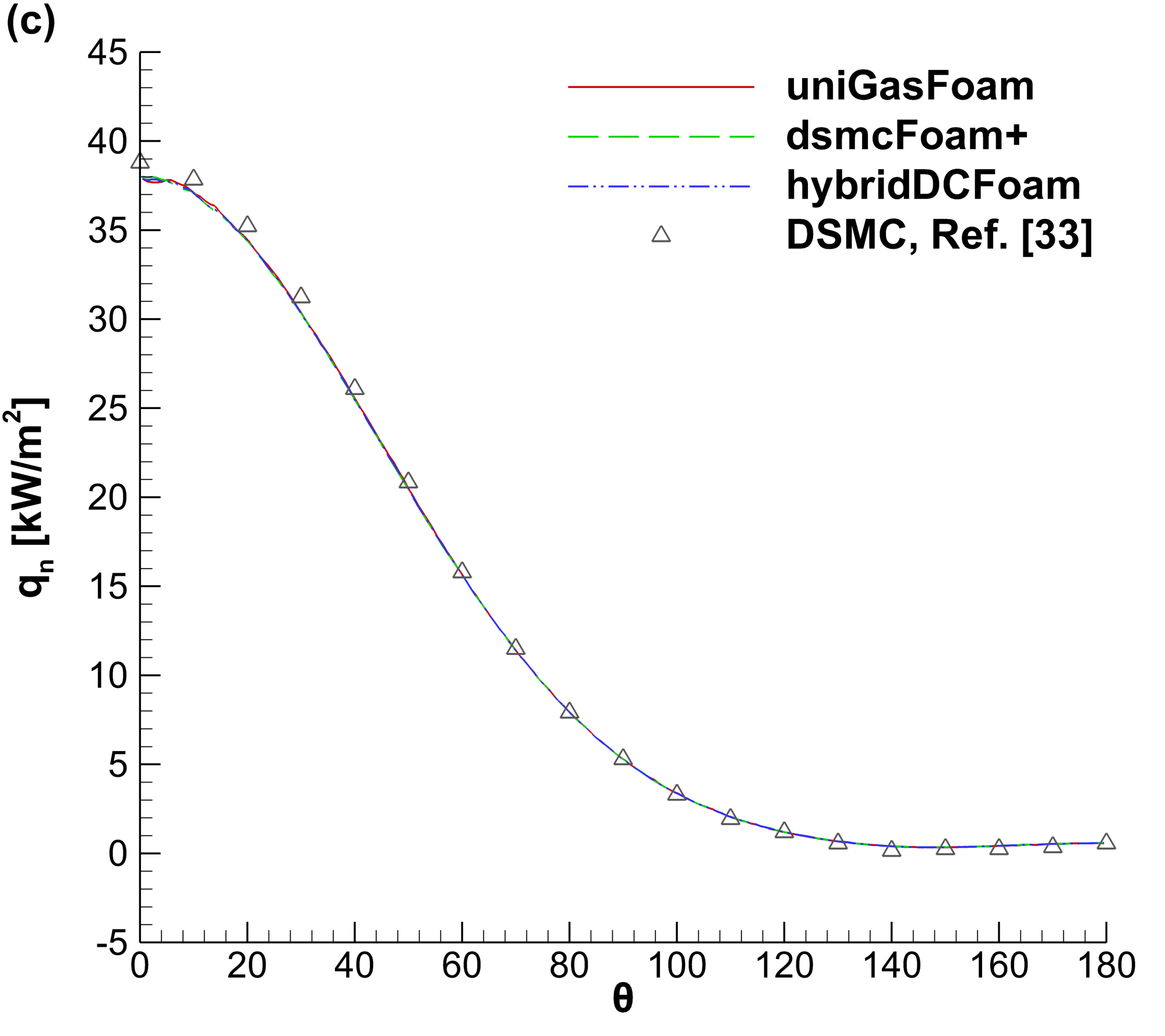}
}
\caption{Cylinder surface (a) normal stress, (b) shear stress and (c) heat flux profiles obtained by uniGasFoam, dsmcFoam+ and hybridDCFoam, along with DSMC results reported in \citep{Lofthouse2007}, for the flow past a cylinder case.}
\label{fig:cylinderPlot}
\end{figure}

A detailed comparison of the cylinder surface normal stress, shear stress, and heat flux predicted by uniGasFoam, dsmcFoam+, and hybridDCFoam solvers is provided in Fig.~\ref{fig:cylinderPlot}. In addition, the DSMC results reported in \citep{Lofthouse2007} are included for validation purposes. It is seen that uniGasFoam produces macroscopic properties in excellent agreement with the dsmcFoam+ and hybridDCFoam solvers, as well as the available literature results, thus validating its accuracy.

The computational cost for the three different methods employed in this study is provided in Table~\ref{tab:cylinderCC}. All presented simulations were run in parallel employing 10 cores. The uniGasFoam and hybridDCFoam solvers employed 26\% and 35\% of computational particles needed in the pure DSMC simulation. Furthermore, a comparable time step was utilized by all approaches, which was constrained by the mean collision time at the cylinder stagnation point. The higher speedup of 3.13 compared to 1.46 can be explained by the fact that uniGasFoam uses fewer computational particles than hybridDCFoam and does not require hybrid iterations.

\begin{table}[h]
\caption{Computational cost of dsmcFoam+, uniGasFoam, and hybridDCFoam for the hypersonic flow past a cylinder case.}
\begin{center}
\begin{tabular}{ l c c c}
\hline
                     & dsmcFoam+ & uniGasFoam & hybridDCFoam \\ \hline
No. particles [mil.] & 12.4      & 3.24       & 4.29  \\ 
Time step [ns]       & 44.3      & 43.8       & 44.7  \\ 
Hybrid iterations        & -         & -          & 2  \\ 
CPU time [hrs]       & 169       & 53.7       & 115  \\ 
Speedup              & -         & 3.13       & 1.46  \\ 
\hline
\end{tabular}
\end{center}
\label{tab:cylinderCC}
\end{table}

\subsection{Flow over plate} \label{Subsect:plate}

\begin{figure}[]
\centerline{
\includegraphics*[width=0.7\textwidth, keepaspectratio=true]{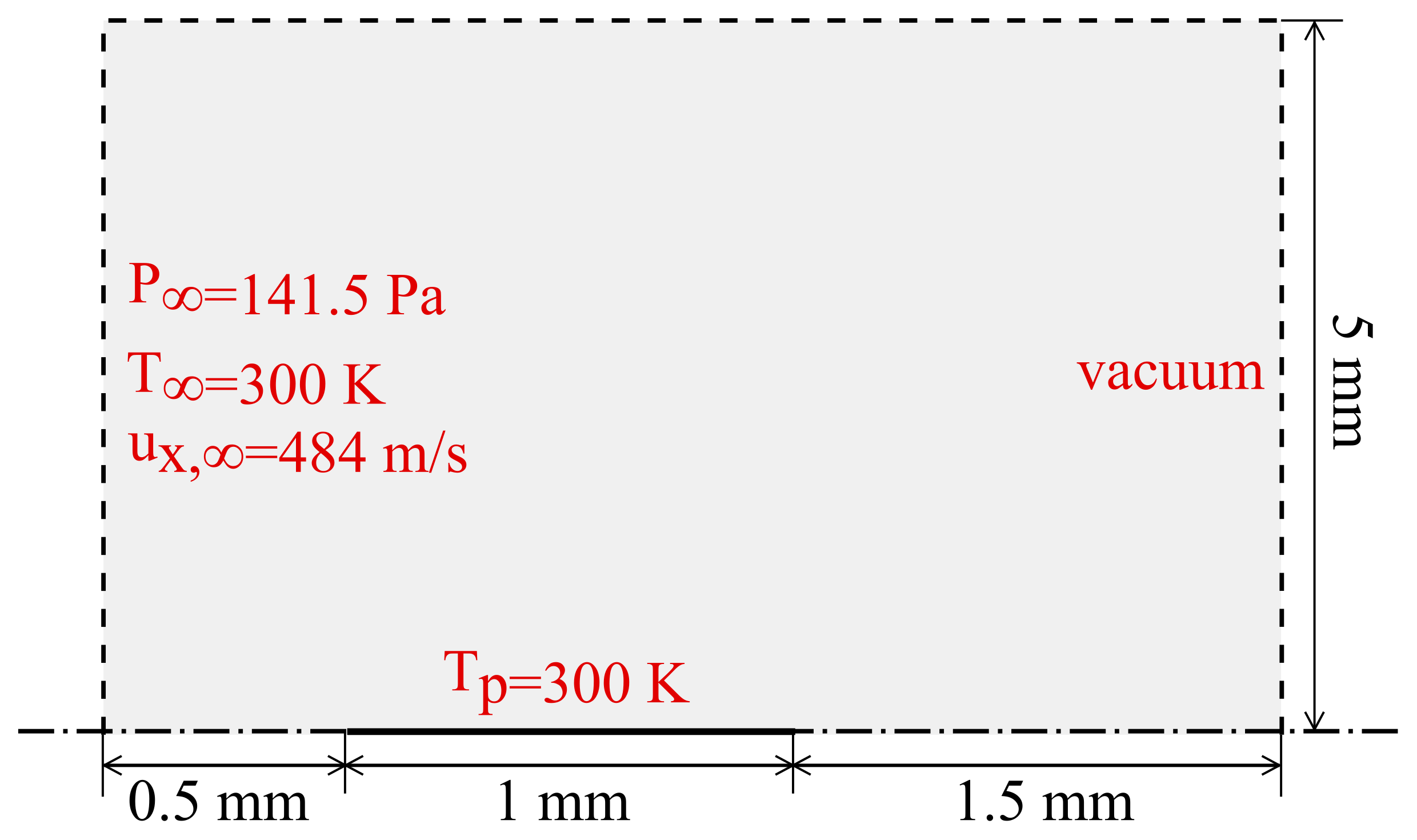}
}
\caption{Configuration of the supersonic flow over a plate benchmark case.}
\label{fig:plateConfig}
\end{figure}

The second benchmark case is the supersonic flow over a flat plate at zero angle of attack. The flow geometry and applied boundary conditions are presented in Fig.~\ref{fig:plateConfig}. Again, in this case, only half of the domain is simulated to reduce the computational cost by taking advantage of the flow symmetry. The modelled gas is argon and the NTC collision scheme with the HS potential is employed according to \citep{Abramov2020}. The specific argon gas properties for the HS potential are taken from \citep{Bird1994}. The plate length is $L$=1 mm, while the left, top, and right boundaries are set 0.5 mm, 5 mm and 1.5 mm away from the plate to minimize their influence on the flow solution. The freestream pressure $P_{\infty}$, temperature $T_{\infty}$, and longitudinal velocity $u_{x,\infty}$ are 141.5 Pa, 300 K, and 484 m/s, which correspond to a Knudsen number calculated based on the plate length of 0.05 and a Mach number of 1.5. The plate is modelled as a purely diffuse isothermal surface with a temperature of $T_p$=300 K. A structured mesh of 225$\times$200 non-uniform hexahedral cells is employed to discretize the flow field. Sampling for steady-state macroscopic quantities commenced at 0.1 ms, and a simulation duration of 4 ms was employed to reduce the statistical scatter.

\begin{figure}[]
\centerline{
\includegraphics*[width=0.6\textwidth, keepaspectratio=true]{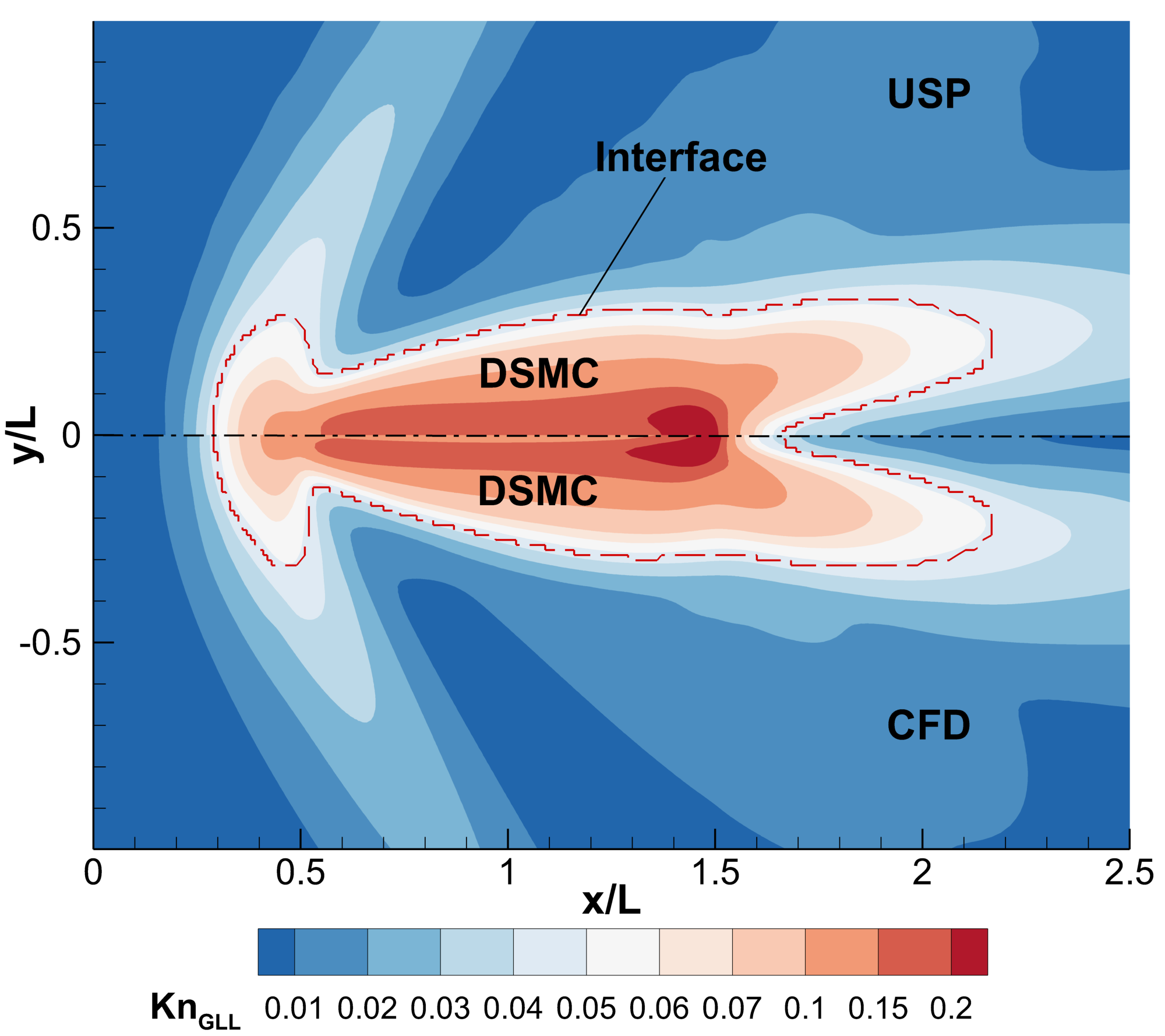}
}
\caption{Maximum gradient local length Knudsen number $\mathrm{Kn}_{GLL}$, along with continuum-rarefied interface predicted by uniGasFoam (top) and hybridDCFoam (bottom), for the flow over a plate case.}
\label{fig:plateDecomp}
\end{figure}

In Fig.~\ref{fig:plateDecomp}, the maximum gradient local length Knudsen number $\mathrm{Kn}_{GLL}$ predicted by uniGasFoam and hybridDCFoam, along with the continuum-rarefied interfaces are provided. It is clearly seen that the continuum-rarefied interface closely tracks the contour $\mathrm{Kn}_{GLL}$=0.05. Additionally, the rarefied area is found around the plate, capturing the leading edge shock, boundary layer, and trailing edge. Here, the velocity gradient is the dominant breakdown criterion, followed by the density and temperature gradients. 

\begin{figure}[]
\centerline{
\includegraphics*[width=0.45\textwidth, keepaspectratio=true]{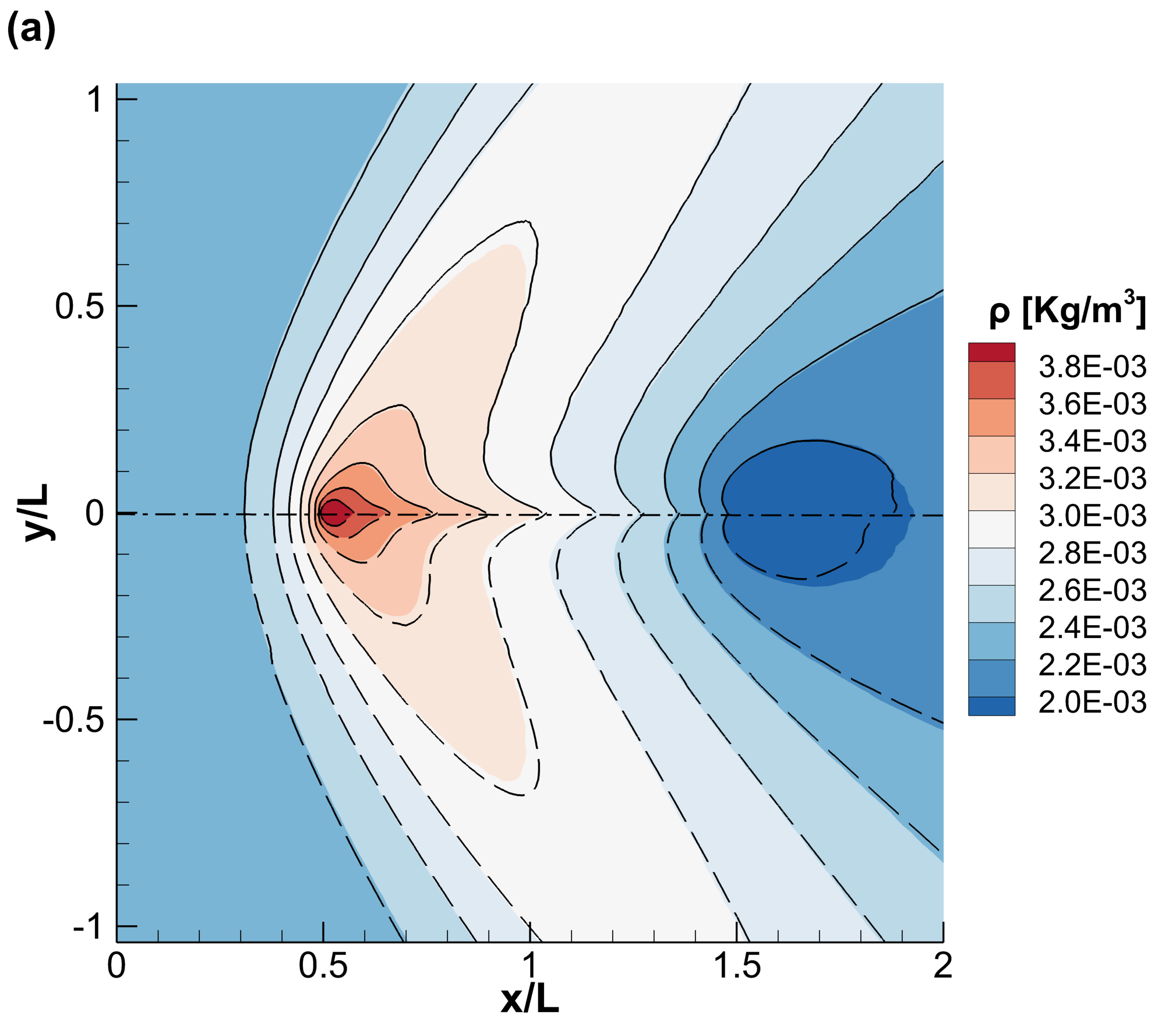}
}
\centerline{
\includegraphics*[width=0.45\textwidth, keepaspectratio=true]{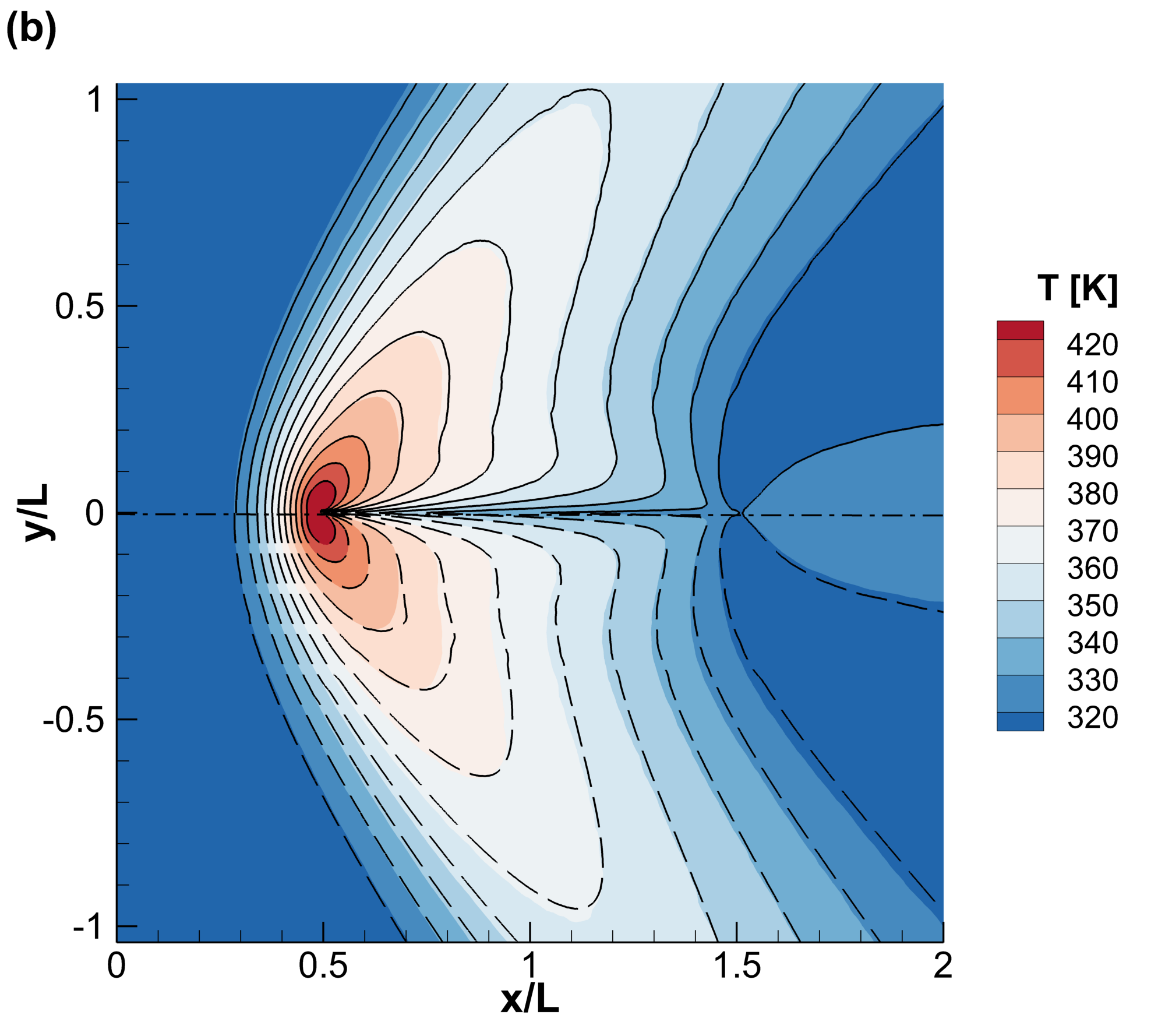}
}
\centerline{
\includegraphics*[width=0.45\textwidth, keepaspectratio=true]{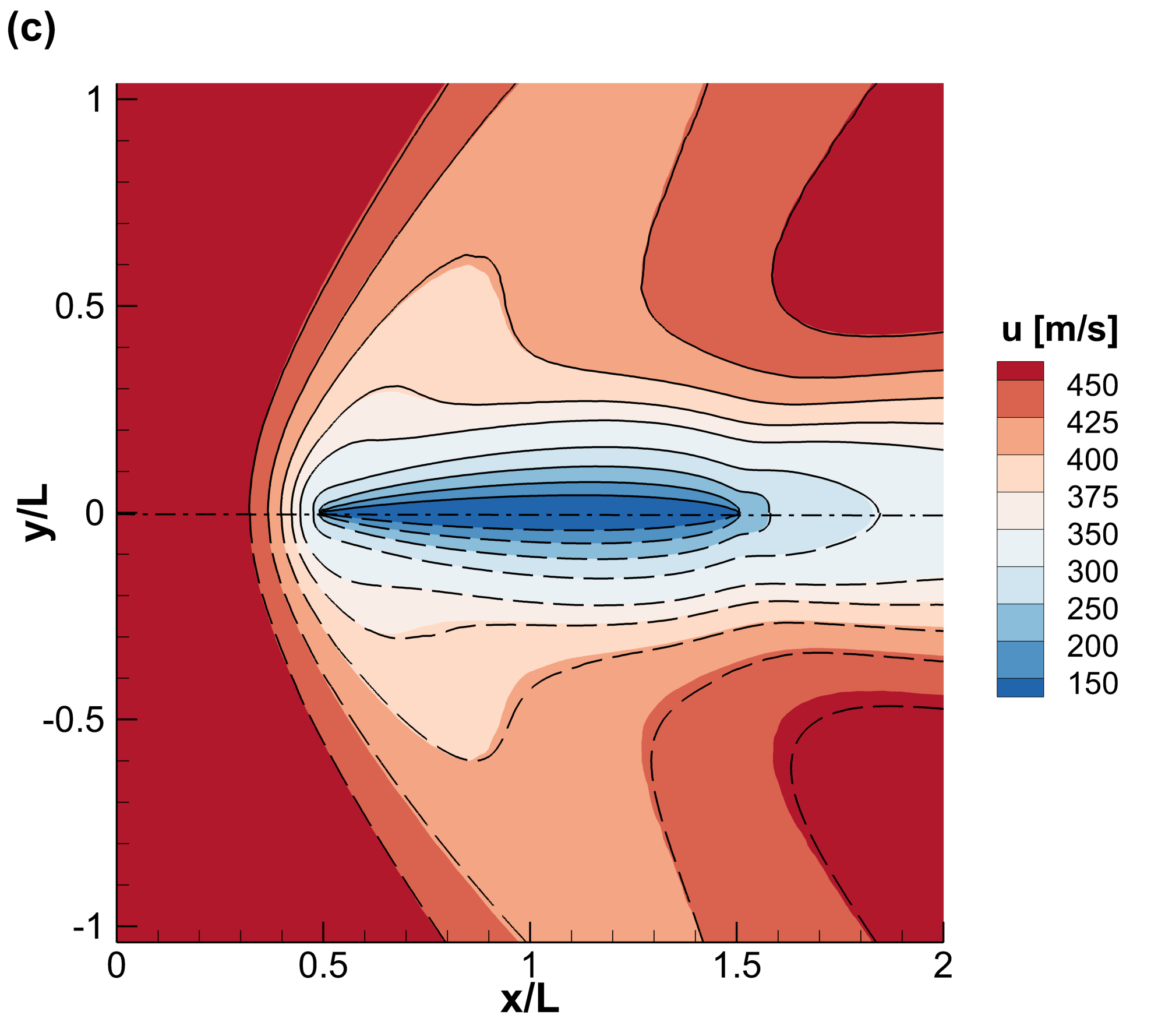}
}
\caption{Comparison of (a) density, (b) temperature, and (c) velocity magnitude obtained by uniGasFoam (solid lines), dsmcFoam+ (colour flood), and hybridDCFoam (dashed lines), for the flow over a plate case.}
\label{fig:plateSlice}
\end{figure}

The density, temperature, and velocity magnitude fields computed by uniGasFoam, dsmcFoam+, and hybridDCFoam, are presented in Fig.~\ref{fig:plateSlice}. An excellent agreement with some minor inconsistencies is observed between uniGasFoam and dsmcFoam+. However, these inconsistencies increase when comparing hybridDCFoam and dsmcFoam+, especially near the trailing edge. It is noted that similar discrepancies between hybrid CFD-DSMC and pure DSMC results have been documented in \citep{Vagishwari2022} for the supersonic flow of argon over a flat plate. Nonetheless, as discussed in the next paragraph, the aforementioned discrepancies do not affect the macroscopic quantities of engineering importance, namely the plate stress and heat flux.

\begin{figure}[]
\centerline{
\includegraphics*[width=0.45\textwidth, keepaspectratio=true]{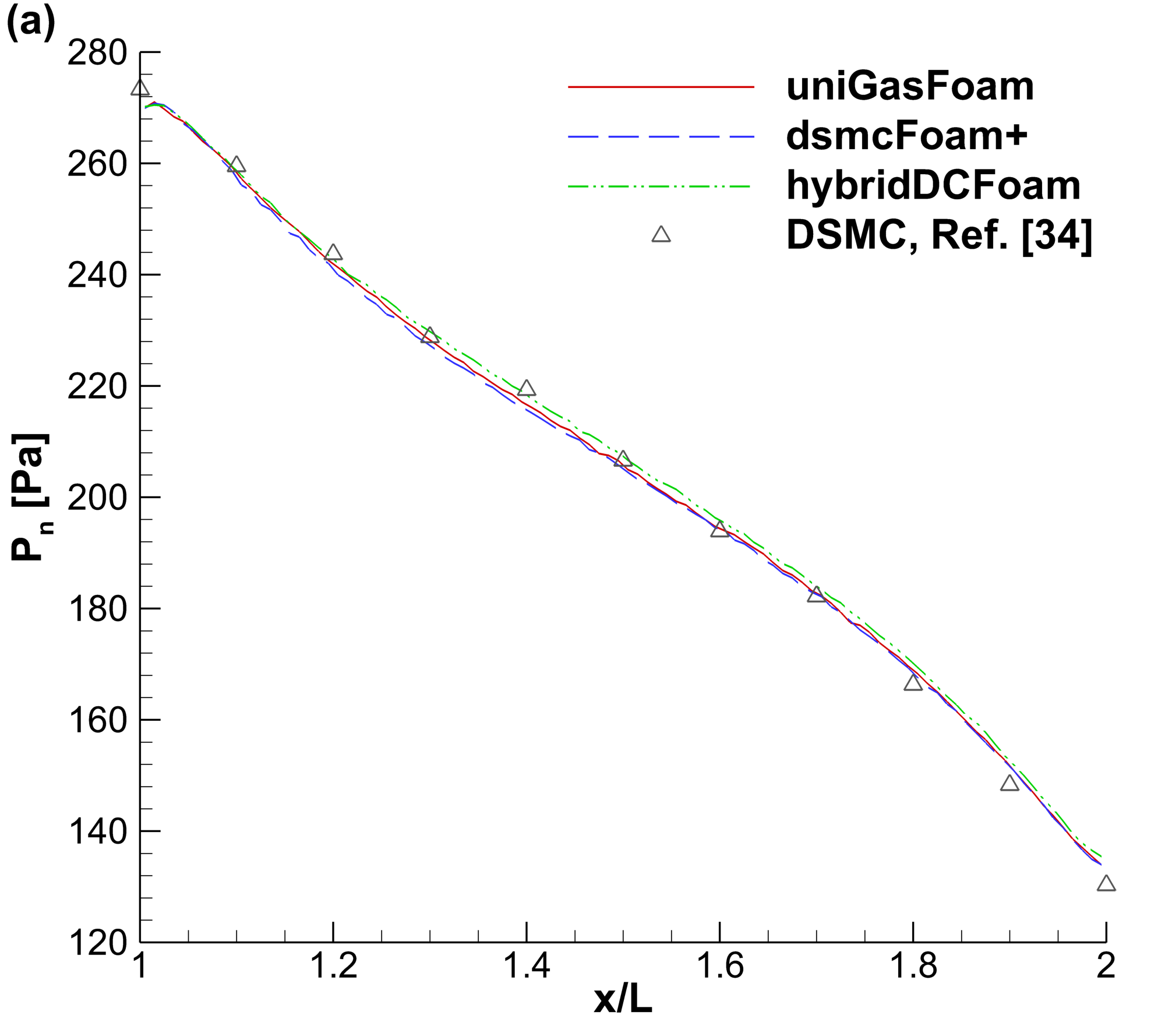}
}
\centerline{
\includegraphics*[width=0.45\textwidth, keepaspectratio=true]{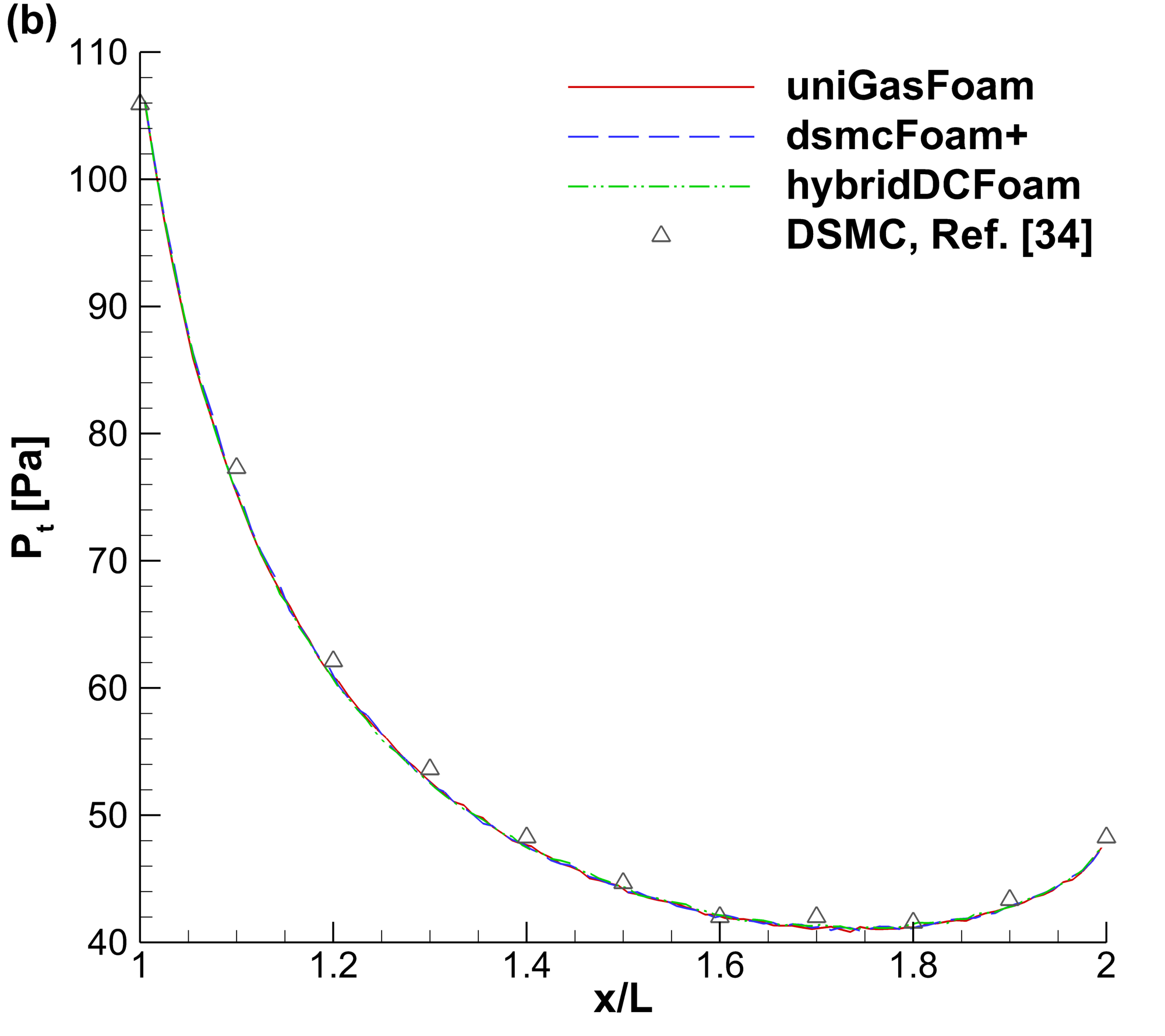}
}
\centerline{
\includegraphics*[width=0.45\textwidth, keepaspectratio=true]{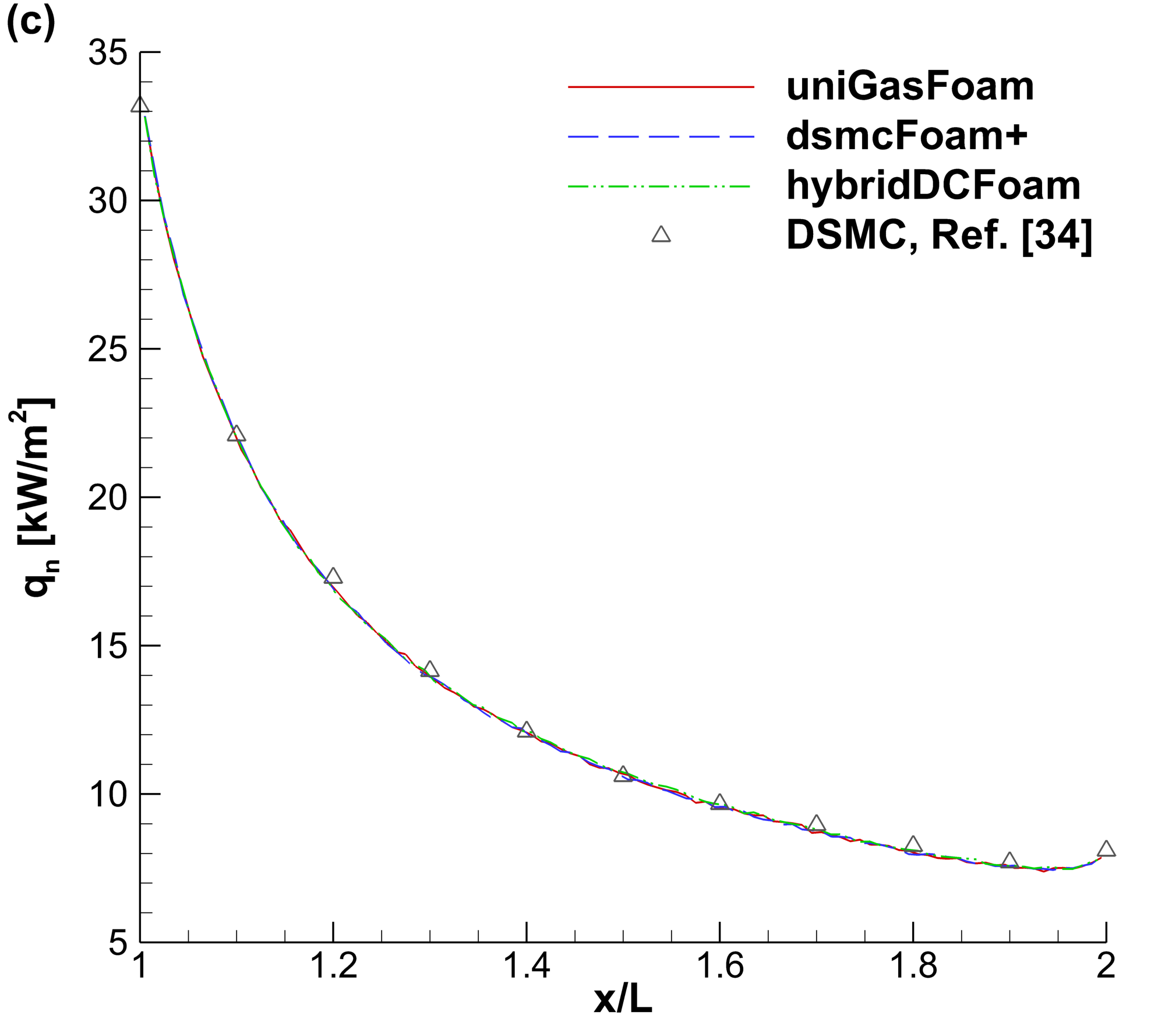}
}
\caption{Plate surface (a) normal stress, (b) shear stress and (c) heat flux profiles obtained by uniGasFoam, dsmcFoam+, and hybridDCFoam, along with DSMC results reported in \citep{Abramov2020}, for the flow over a plate case.}
\label{fig:platePlot}
\end{figure}

In Fig.~\ref{fig:platePlot} a detailed comparison of the plate surface normal stress, shear stress, and heat flux obtained by uniGasFoam, dsmcFoam+, and hybridDCFoam is presented. Additionally, pure DSMC simulation results reported in \citep{Abramov2020} are included for benchmarking. The uniGasFoam solver provides an excellent agreement with dsmcFoam+, hybridDCFoam, and the reported literature results, hence validating the numerical accuracy of the USP-DSMC scheme.

In Table~\ref{tab:plateCC} the computational cost of the dsmcFoam+, uniGasFoam, and hybridDCFoam simulations is provided. All presented simulations were executed in parallel and utilized 5 cores. In this scenario, the uniGasFoam and hybridDCFoam solvers used 62\% and 7.7\% of particles required for a pure dsmcFoam+ simulation, respectively. Moreover, all methods used a comparable time step, which was determined by the mean collision time at the leading edge of the plate. The hybridDCFoam solver achieved a speedup of 3.71 compared to dsmcFoam+ overperforming uniGasFoam, which achieved a speedup of 1.26. This outcome is justified by the fact that the rarefied region is much smaller than the continuum region, enabling hybridDCFoam to use a considerably smaller number of computational particles.

\begin{table}[h]
\caption{Computational cost required by dsmcFoam+, uniGasFoam and hybridDCFoam for the supersonic flow over a plate case.}
\begin{center}
\begin{tabular}{ l c c c}
\hline
                      & dsmcFoam+ & uniGasFoam & hybridDCFoam \\ \hline
No. particles [mil.]  & 1.90      & 1.19       & 0.147 \\ 
Time step [ns]        & 8.51      & 8.21       & 8.23  \\ 
Hybrid iterations         & -         & -          & 3     \\ 
CPU time [hrs]        & 159       & 126        & 42.8  \\ 
Speedup               & -         & 1.26       & 3.71  \\ 
\hline
\end{tabular}
\end{center}
\label{tab:plateCC}
\end{table}

\subsection{Nozzle plume impingement} \label{Subsect:nozzle}

\begin{figure}[]
\centerline{
\includegraphics*[width=0.8\textwidth, keepaspectratio=true]{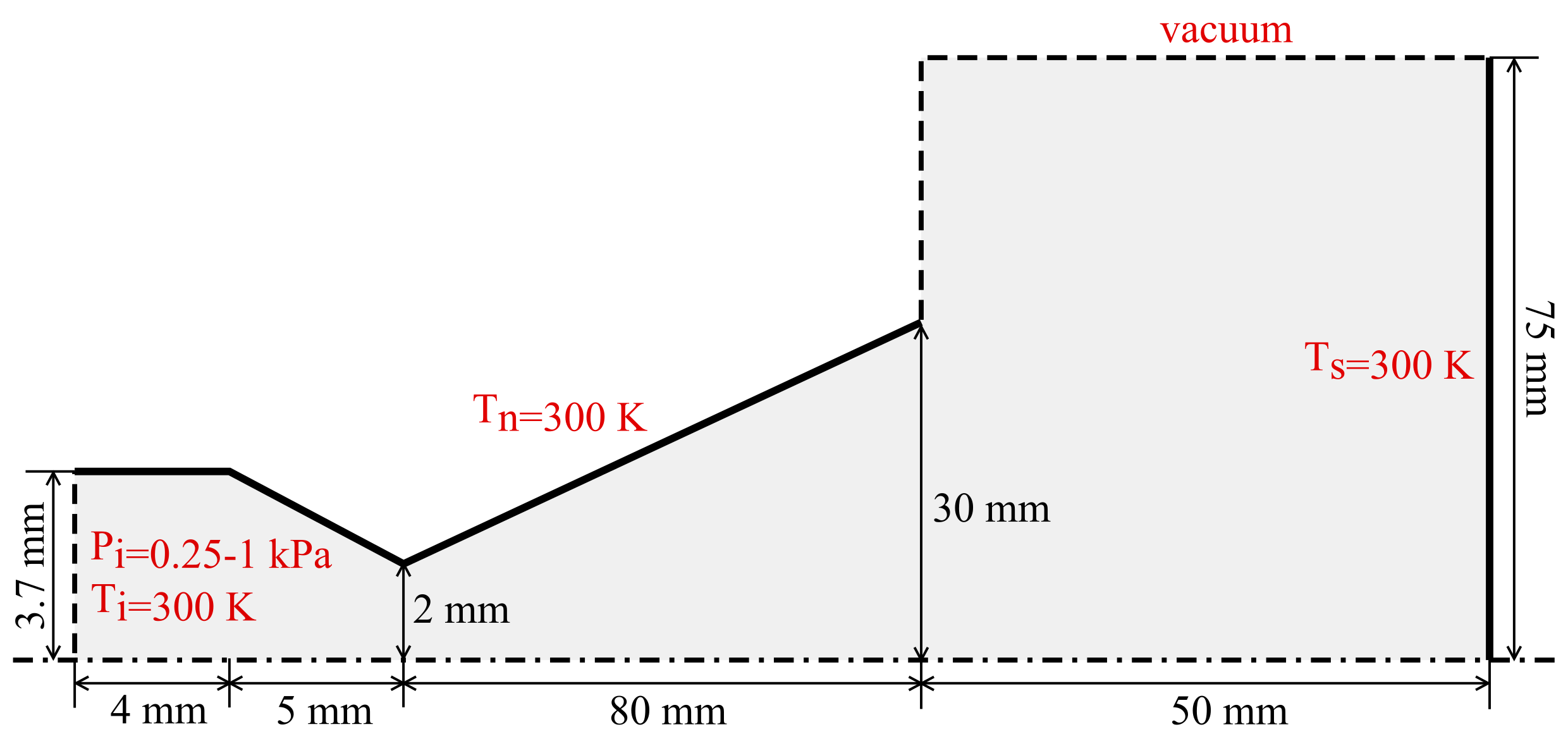}
}
\caption{Configuration of the nozzle plume impingement benchmark case.}
\label{fig:nozzleConfig}
\end{figure}

In the next benchmark case, the plume impingement of a conical nozzle is investigated. The exact dimensions of the flow configuration, along with the applied boundary conditions, are illustrated in Fig.~\ref{fig:nozzleConfig}. The stagnation chamber contains argon at a temperature of $T_i$=300 K, while different inlet pressures $P_i$=[0.25,0.5,0.75,1] Pa are considered. The argon gas expands through the nozzle into vacuum, impinging on the solid surface, which is located downstream from the nozzle outlet. In all cases studied, the flow is choked at the nozzle throat reaching Mach 1. The nozzle walls and impingement surface are modelled as fully diffuse and are kept at a constant temperature of $T_n$=$T_s$=300 K. The NTC collision scheme with the VHS potential is employed to model the argon gas and the specific gas properties are taken from \citep{Bird1994}. It is noted that, due to the flow axial symmetry, only a 5$^\circ$ wedge is simulated to save computational resources. The computational flow domain was discretized using a structured mesh consisting of 78,000 non-uniform cells. In all cases, steady-state sampling started at 1 ms, and a simulation duration of 3 ms was employed to reduce the statistical scatter.

\begin{figure}[]
\centerline{
\includegraphics*[width=0.6\textwidth, keepaspectratio=true]{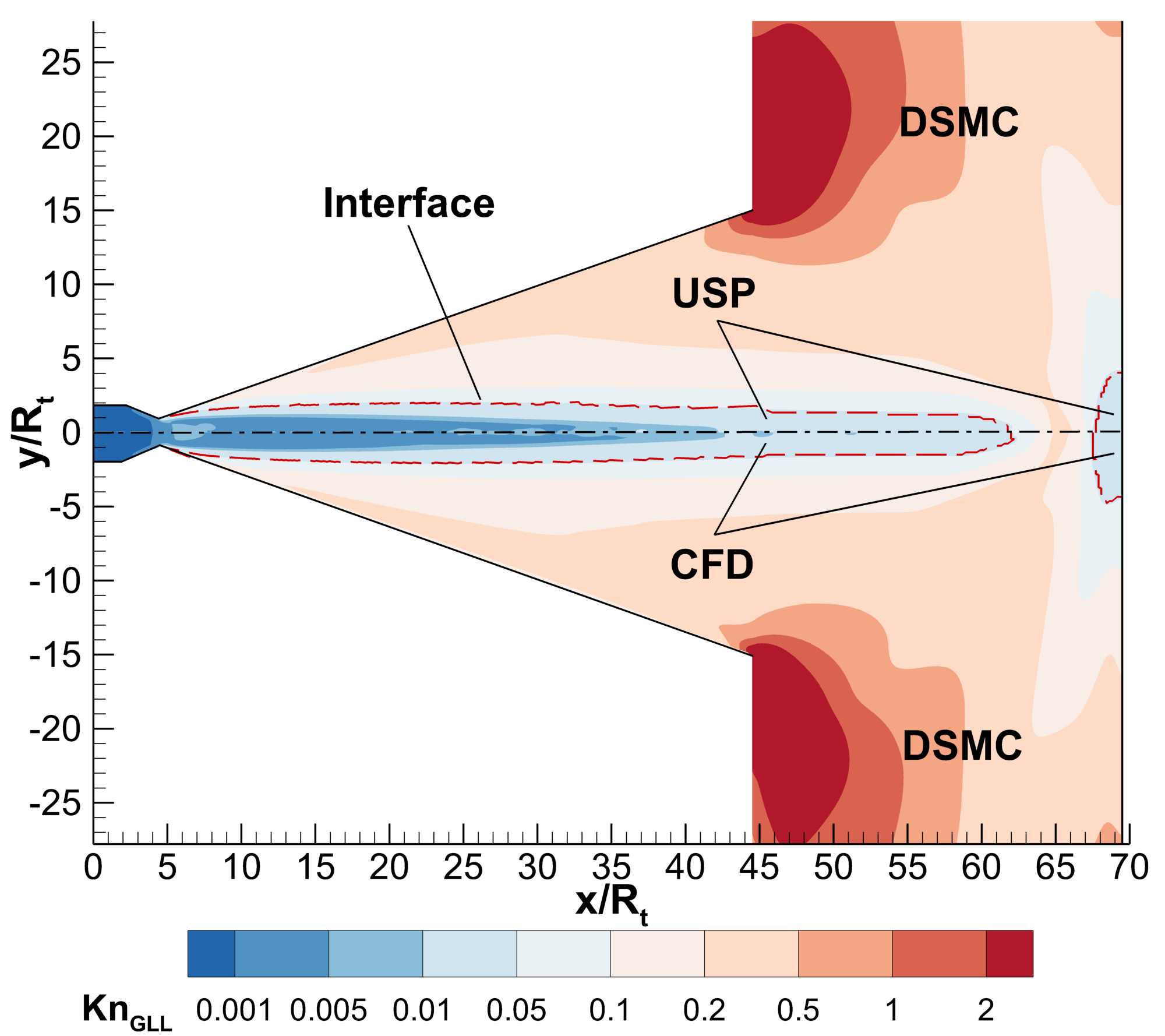}
}
\caption{Maximum gradient local length Knudsen number $\mathrm{Kn}_{GLL}$, along with continuum-rarefied interface predicted by uniGasFoam (top) and hybridDCFoam (bottom), for the nozzle plume impingement case; $P_i$=1 $kPa$.}
\label{fig:nozzleSliceKn}
\end{figure}

The maximum gradient local length Knudsen number $\mathrm{Kn}_{GLL}$ obtained by uniGasFoam and hybridDCFoam, along with the continuum-rarefied interfaces, in the case of $P_i$=1 kPa, are presented in Fig.~\ref{fig:nozzleSliceKn}. The continuum region is predominantly located within the converging section of the nozzle due to the high pressure and small flow gradients in this area. In addition, a localized continuum region is observed in the centre of the downstream surface due to the sudden pressure increase from the impinging flow. Conversely, the rarefied region is concentrated in the diverging section of the nozzle due to the rapid gas expansion downstream of the nozzle throat.  In this case, the density gradient criterion is the dominant one, followed by the temperature and velocity gradient criteria. Furthermore, the continuum-rarefied interface strictly follows the gradient local length Knudsen splitting value of 0.05, in both hybrid solvers. 

\begin{figure}[]
\centerline{
\includegraphics*[width=0.45\textwidth, keepaspectratio=true]{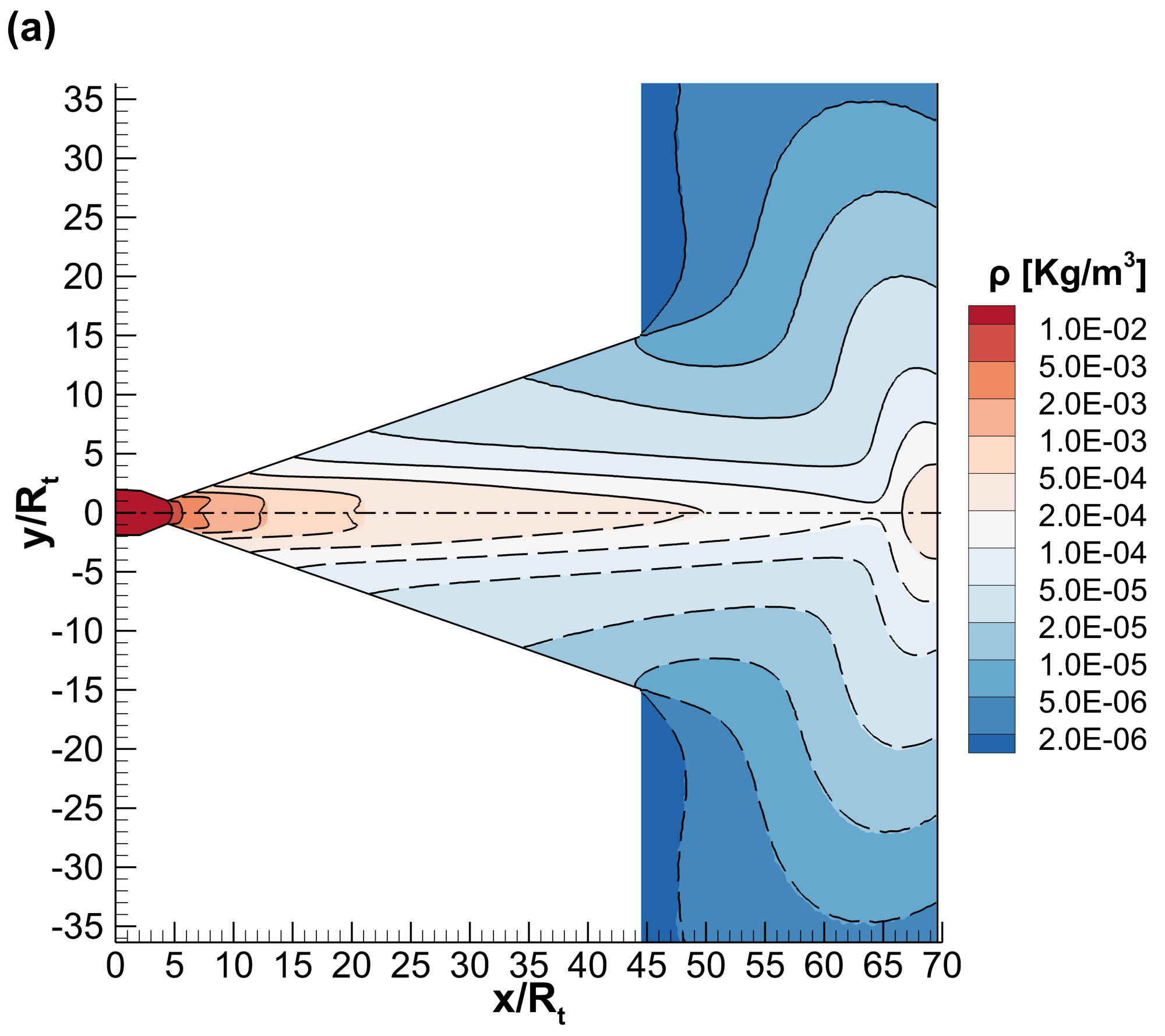}
}
\centerline{
\includegraphics*[width=0.45\textwidth, keepaspectratio=true]{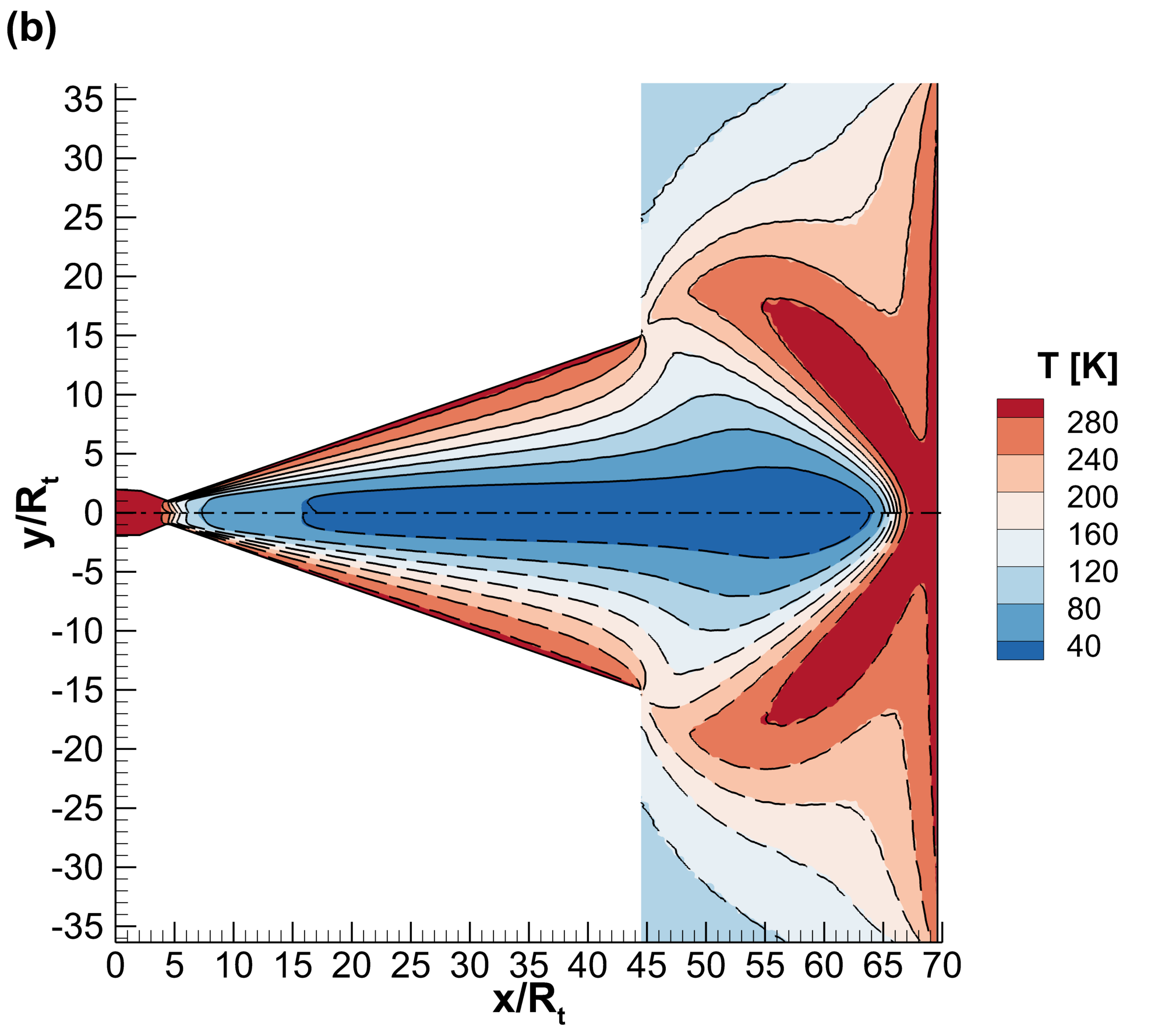}
}
\centerline{
\includegraphics*[width=0.45\textwidth, keepaspectratio=true]{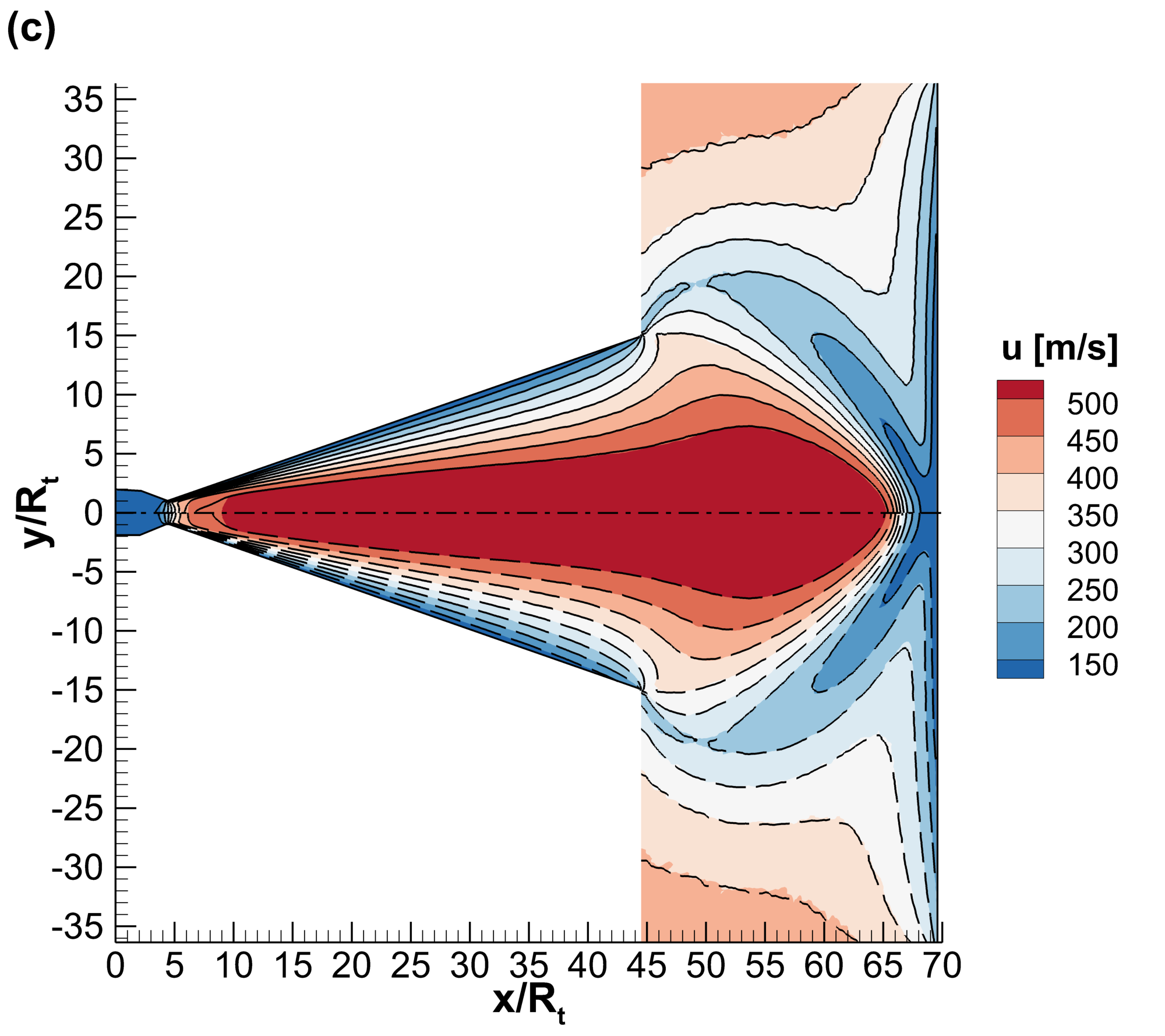}
}
\caption{Comparison of (a) density, (b) temperature and, (c) velocity magnitude obtained by uniGasFoam (solid lines), dsmcFoam+ (colour flood) and, hybridDCFoam (dashed lines), for the nozzle plume impingement case; $P_i$=1 $kPa$.}
\label{fig:nozzleSlice}
\end{figure}

In Fig.~\ref{fig:nozzleSlice} contour plots of the density, temperature, and velocity magnitude obtained by uniGasFoam, dsmcFoam+, and hybridDCFoam, in the case of $P_i$=1 kPa are provided. In the diverging section of the nozzle, the gas undergoes rapid expansion, resulting in a decrease in both density and temperature while the velocity increases. Further downstream, as the gas plume impinges on the target surface, it decelerates, leading to an increase in density and temperature and a corresponding decrease in velocity. In addition, both the uniGasFoam and hybridDCFoam solvers are in excellent agreement with dsmcFoam+, with some small discrepancies observed near the flow axis. It is noted that although only the case of $P_i$=1 kPa was discussed here, all remarks still hold for the lower inlet pressure cases.

\begin{figure}[]
\centerline{
{\includegraphics*[width=0.45\textwidth, keepaspectratio=true]{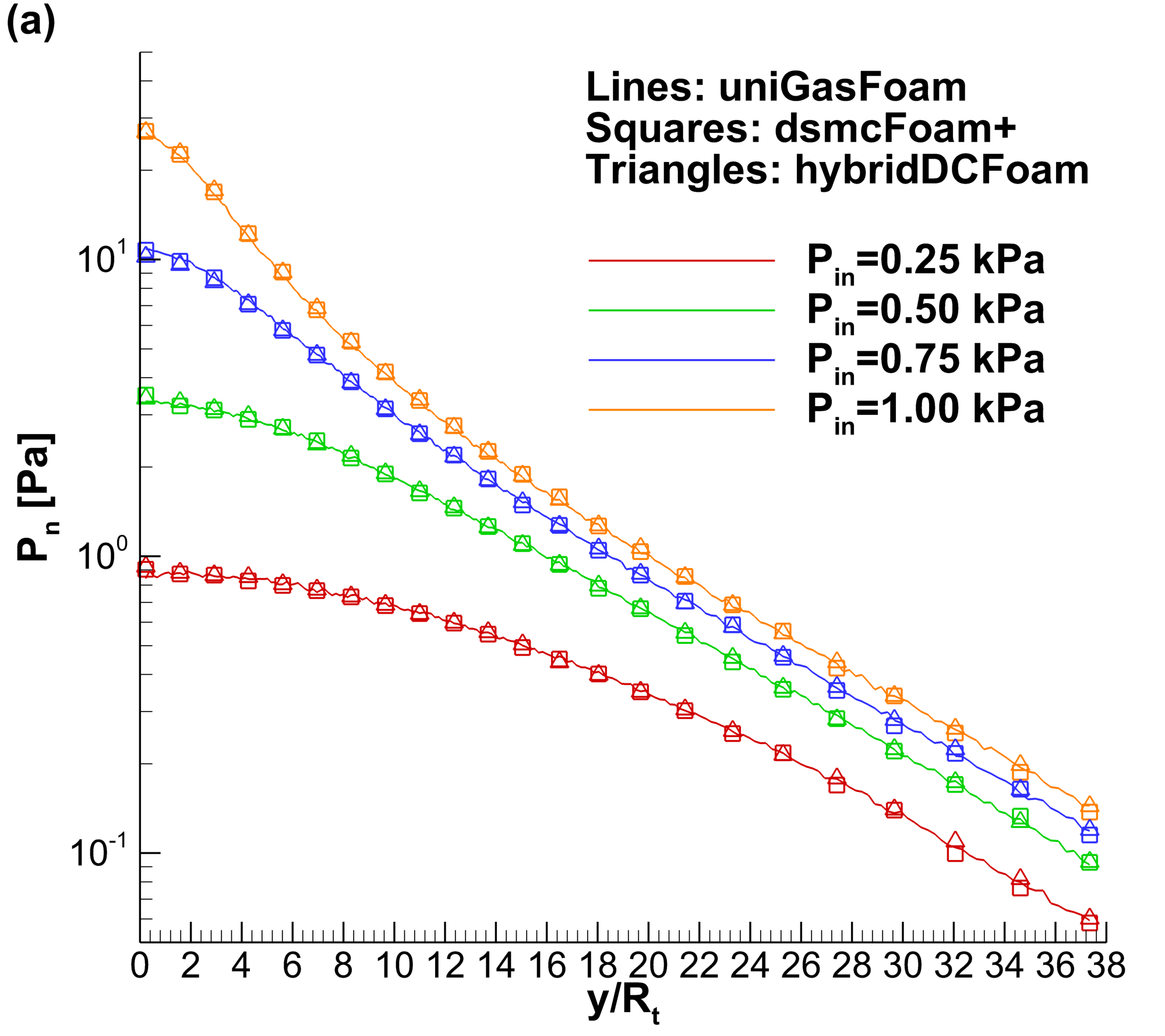}
\includegraphics*[width=0.45\textwidth, keepaspectratio=true]{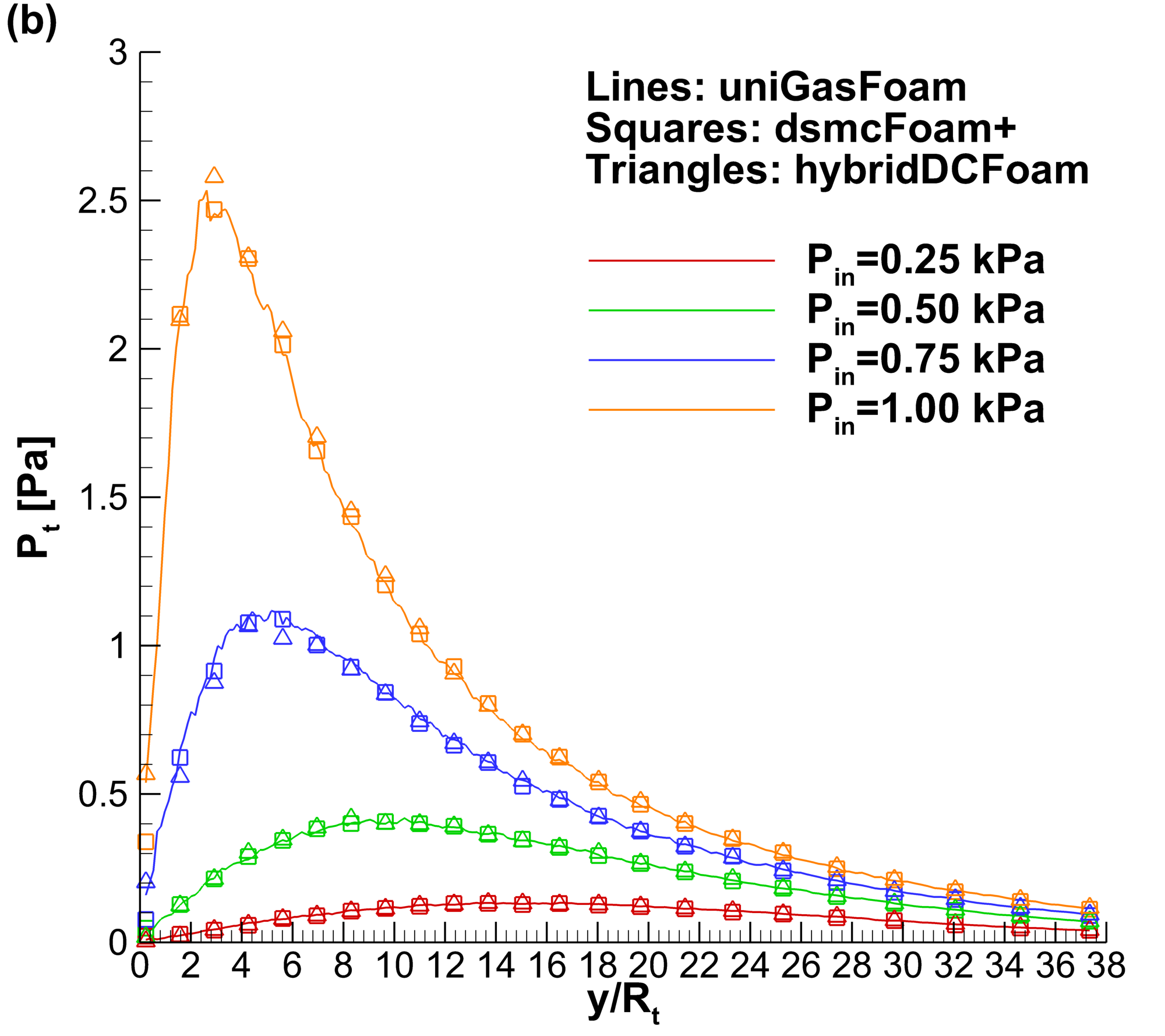}}
}
\caption{Impingement surface (a) normal stress and (b) shear stress profiles obtained by uniGasFoam, dsmcFoam+, and hybridDCFoam for the nozzle plume impingement case.}
\label{fig:nozzlePlot}
\end{figure}

In Fig.~\ref{fig:nozzlePlot} a detailed comparison of the impingement surface normal stress and shear stress predicted by uniGasFoam, dsmcFoam+, and hybridDCFoam is shown. As expected as the nozzle inlet pressure is increased, both normal and shear stresses are increased, while the maximum shear stress location moves closer to the symmetry axis. It is clearly seen that uniGasFoam is in excellent agreement with dsmcFoam+ and hybridDCFoam in the whole nozzle inlet pressure range investigated, thus validating its ability to provide accurate predictions for these multiscale problems that contain large changes in $\mathrm{Kn}$ in one simulation domain. In this case, the $\mathrm{Kn}$ changes three orders of magnitude, from continuum to late transition, as shown in Fig.~\ref{fig:nozzleSliceKn}.

\begin{figure}[]
\centerline{
\includegraphics*[width=0.6\textwidth, keepaspectratio=true]{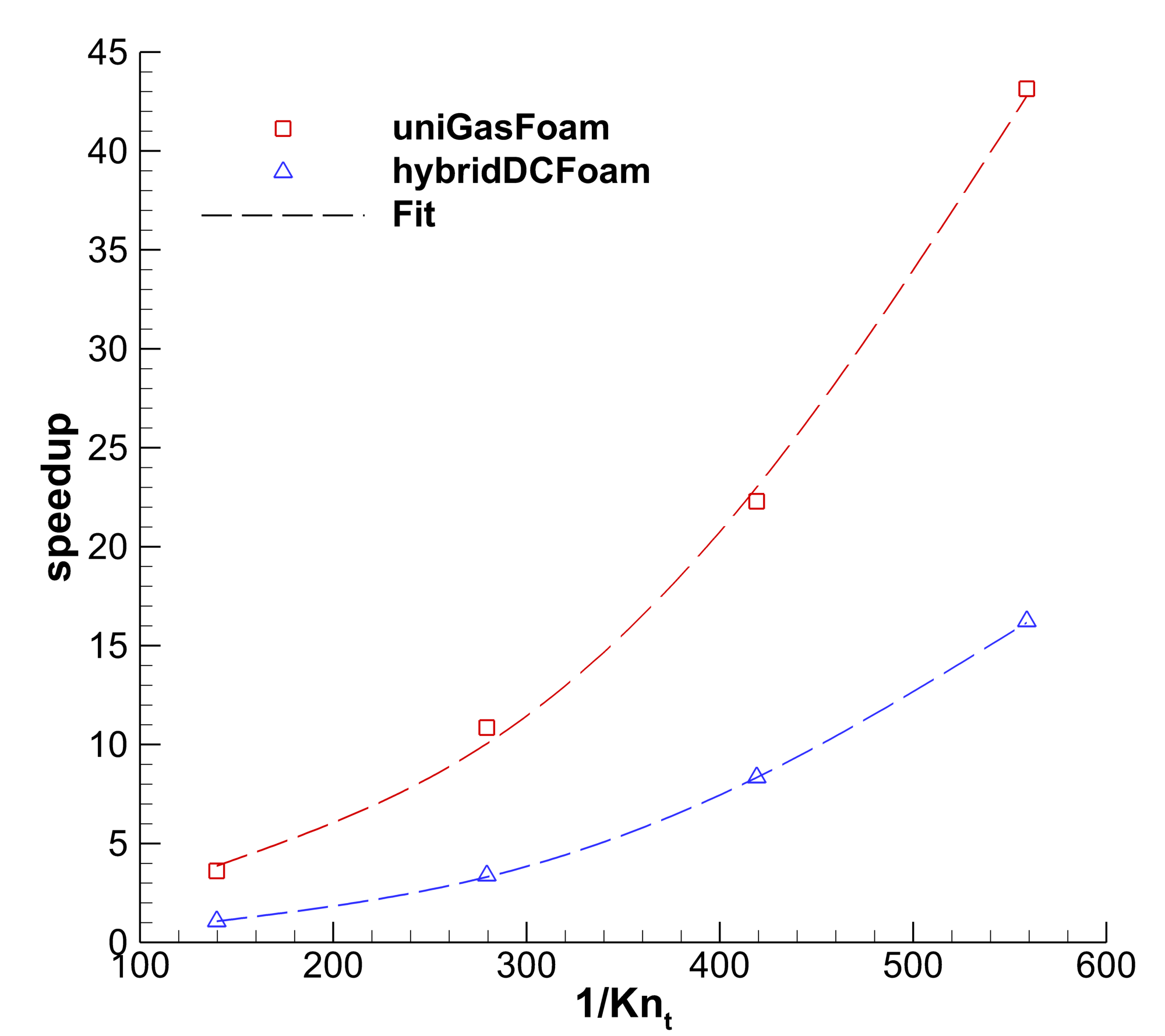}
}
\caption{Speedup obtained by uniGasFoam and hybridDCFoam simulations over dsmcFoam+ with respect to the inverse nozzle throat Knudsen number for the nozzle plume impingement benchmark case.}
\label{fig:nozzlePlotSpeedup}
\end{figure}

The computational cost of the dsmcFoam+, uniGasFoam, and hybridDCFoam solvers is presented in Table~\ref{tab:nozzleCC}. All presented simulations were performed in parallel and employed 25 cores. It is evident that as the nozzle inlet pressure increases, the pure DSMC computational effort rapidly increases due to constraints on cell size and time step. In contrast, the number of particles and time step used by uniGasFoam and hybridDCFoam remain relatively constant, resulting in only a marginal increase in overall computational cost for the higher inlet pressures. The hybridDCFoam solver employs a slightly smaller time step than uniGasFoam, as it must resolve the mean collision time within the buffer region, which extends further upstream than the rarefied region. Conversely, uniGasFoam requires a larger number of particles to resolve the entire flow domain. However, because it does not require hybrid iterations, uniGasFoam ultimately outperforms hybridDCFoam in terms of the overall computational cost. The speedup achieved by uniGasFoam and hybridDCFoam over dsmcFoam+ with respect to the inverse throat Knudsen number is illustrated in Fig.~\ref{fig:nozzlePlotSpeedup}. In both cases, a polynomial increase in speedup is observed as the throat Knudsen number decreases. However, as previously discussed, uniGasFoam consistently outperforms hybridDCFoam, providing approximately 3 times more speedup. These approaches become more efficient as the Knudsen number decreases and the flow becomes more multiscale. Extrapolating from the presented results, it is evident that at some point, the implementation of pure DSMC becomes computationally infeasible, making hybrid methods the only viable solution for addressing these flows.


\begin{table}[h]
\caption{Computational cost required by dsmcFoam+, uniGasFoam, and hybridDCFoam for the nozzle plume impingement case.}
\begin{center}
\begin{tabular}{ l c c c c}

\hline
Inlet pressure [kPa] & 0.25 & 0.5 & 0.75 & 1 \\
\hline
\multicolumn{5}{c}{dsmcFoam+}\\ 
\hline
No. particles [mil.] &4.93 & 10.9 & 20.4 & 32.1 \\ 
Time step [ns] & 7.59 & 4.42 & 3.13 & 2.42 \\
CPU time [hrs] & 69.1 & 253 & 633 & 1320 \\
\hline
\multicolumn{5}{c}{uniGasFoam}\\
\hline
No. particles [mil.] & 2.50 & 2.56 & 2.65 & 2.79 \\ 
Time step [ns] & 18.6 & 14.8 & 12.9 & 13.1 \\
CPU time [hrs] & 19.1 & 23.3 & 28.4 & 30.6 \\
Speedup & 3.62 & 10.9 & 22.3 & 43.1 \\ 
\hline
\multicolumn{5}{c}{hybridDCFoam}\\ 
\hline
No. particles [mil.]  & 1.82 &1.83 & 1.85 & 1.87 \\ 
Time step [ns] & 13.5 & 11.7 & 11.7 & 11.0 \\
Hybrid iterations & 3 & 3 & 3 & 3 \\
CPU time [hrs] & 64.3 & 74.8 & 75.9 & 81.3 \\
Speedup & 1.07 & 3.38 & 8.34 & 16.2 \\ 
\hline
\end{tabular}
\end{center}
\label{tab:nozzleCC}
\end{table}

\subsection{Transient gas expansion into vacuum} \label{Subsect:expansion}

\begin{figure}[]
\centerline{
\includegraphics*[width=0.8\textwidth, keepaspectratio=true]{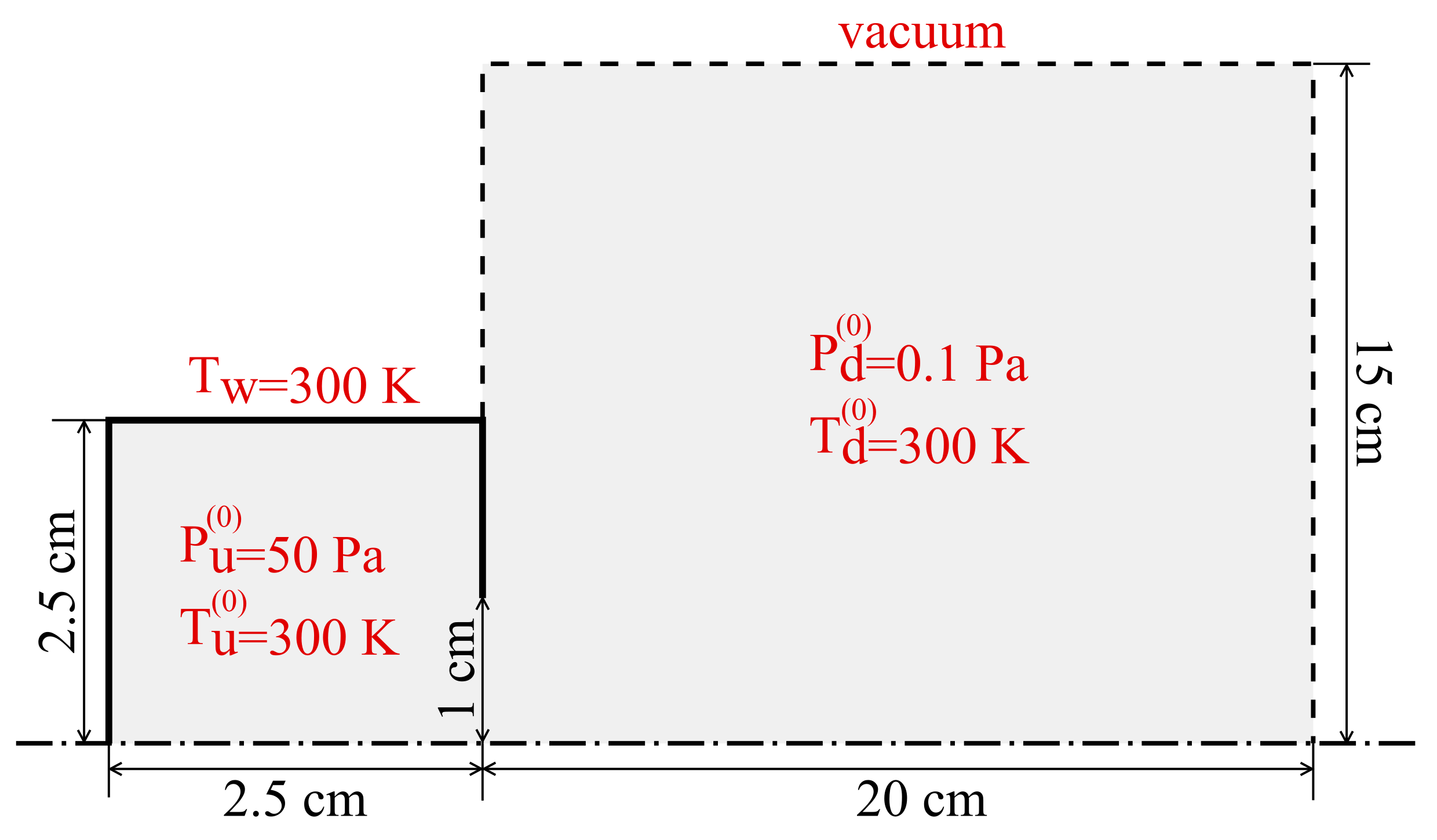}
}
\caption{Configuration of the transient expansion into vacuum benchmark case.}
\label{fig:expansionConfig}
\end{figure}

The final benchmark considered in this work is a transient expansion into vacuum. The exact dimensions of the flow domain, along with the applied boundary conditions are illustrated in Fig.~\ref{fig:expansionConfig}. Initially, the cylindrical upstream vessel is sealed, containing argon gas at an initial pressure $P_u^{(0)}$=50 Pa and a temperature of $P_u^{(0)}$=300 K. At $t$=0 s the vessel orifice is opened and the gas starts to expand into vacuum. The vessel walls are modelled as a fully diffuse isothermal surface with a constant temperature of $T_w$=300 K, while the initial vacuum pressure $P_d^{(0)}$ and temperature $T_d^{(0)}$ are set to 0.1 Pa and 300 K, respectively. The VHS potential is used to model the argon gas, with the specific gas properties taken from \citep{Bird1994}, while the NTC collision scheme is employed. Exploiting the flow axial symmetry, only a 5$^\circ$ wedge is simulated to reduce the required computational effort. It is noted that, for this case, only the uniGasFoam and dsmcFoam+ solvers are employed, since hybridDCFoam is limited only to steady-state cases. A structured mesh of 17,500 non-uniform cells was employed to discretize the flow domain. In order to sufficiently reduce the statistical noise of the transient results, the desired number of particles per sub-cell was set to 200.

\begin{figure}[]
\centerline{
{\includegraphics*[width=0.45\textwidth, keepaspectratio=true]{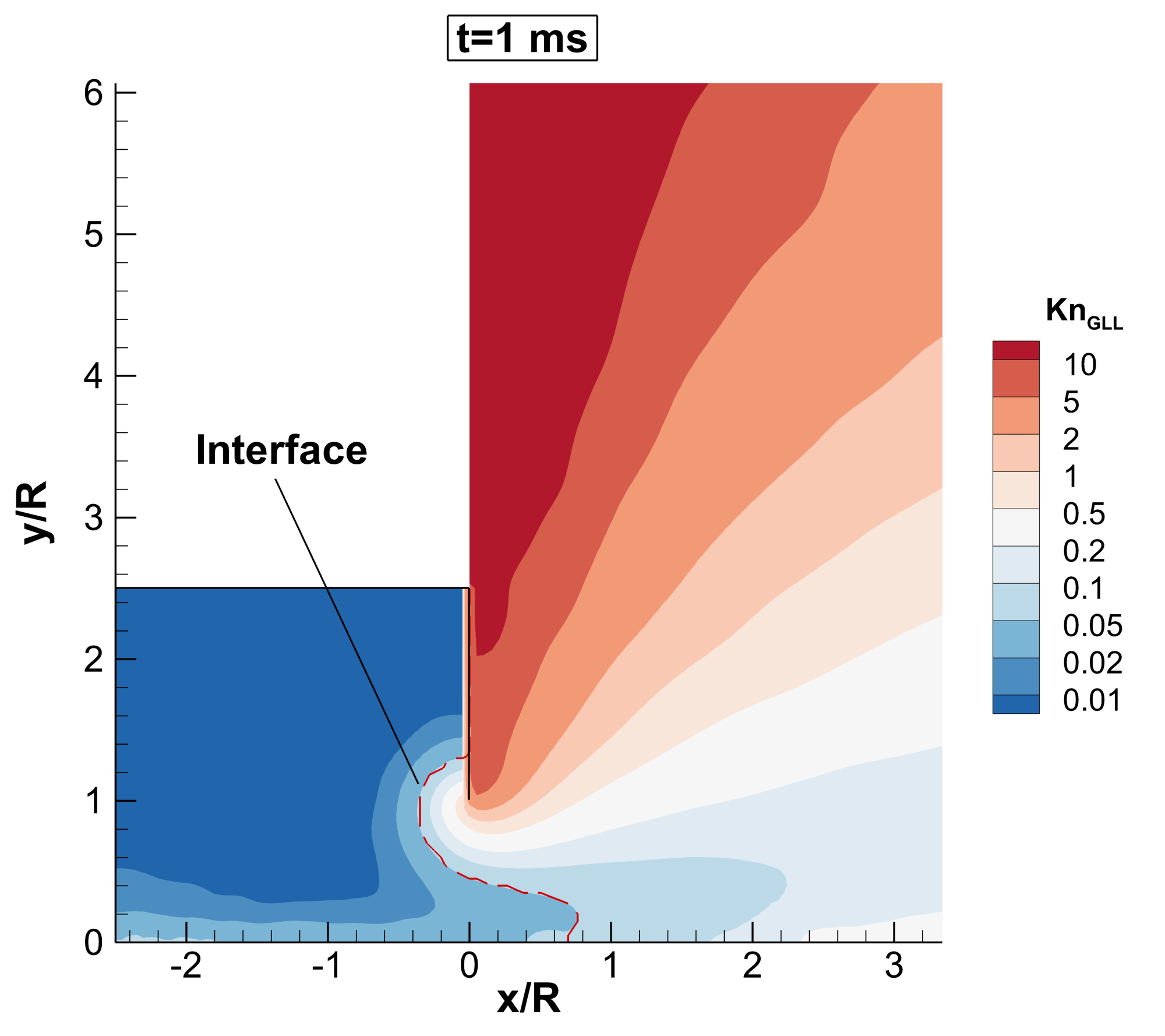}
\includegraphics*[width=0.45\textwidth, keepaspectratio=true]{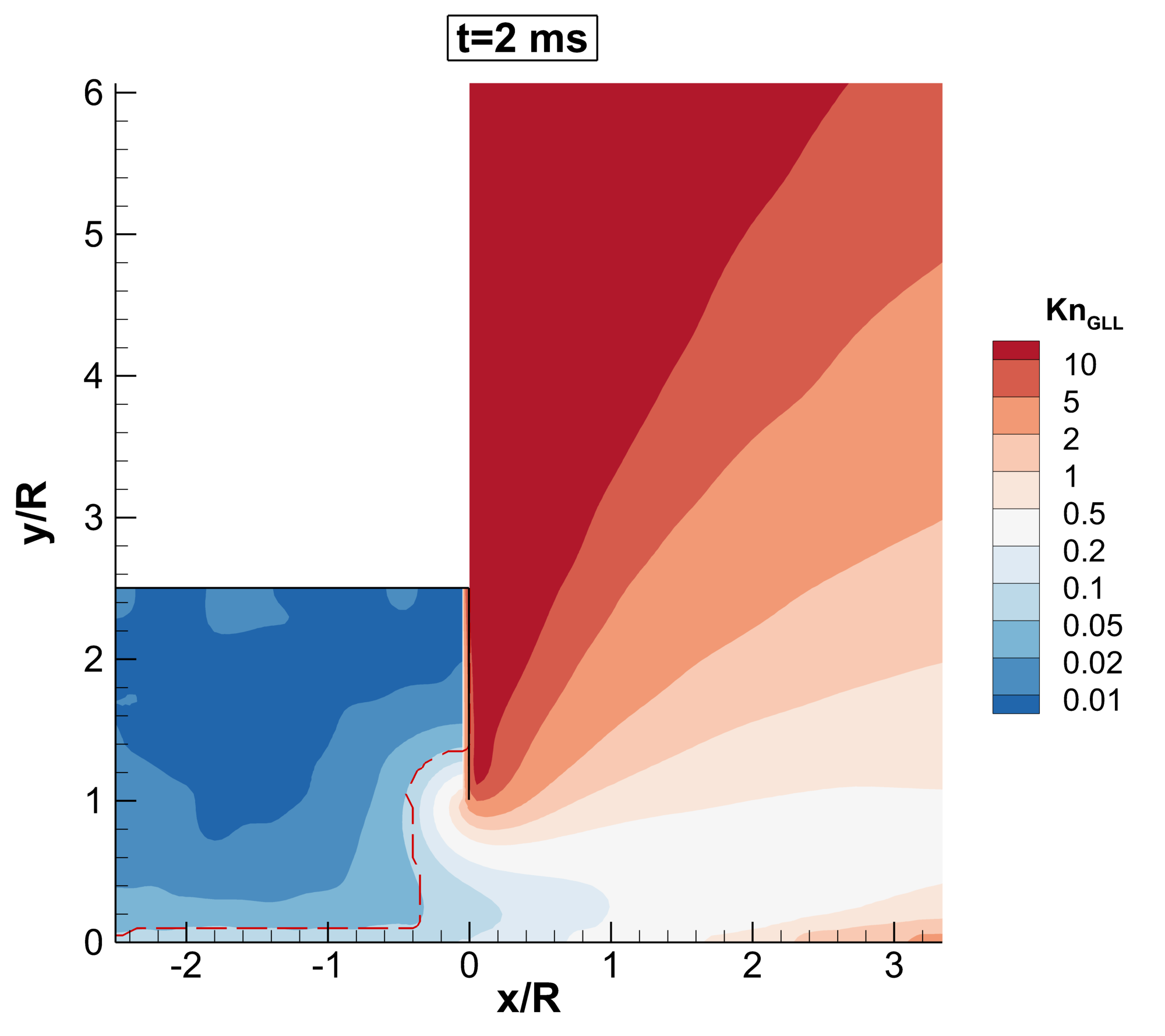}}
}
\centerline{
{\includegraphics*[width=0.45\textwidth, keepaspectratio=true]{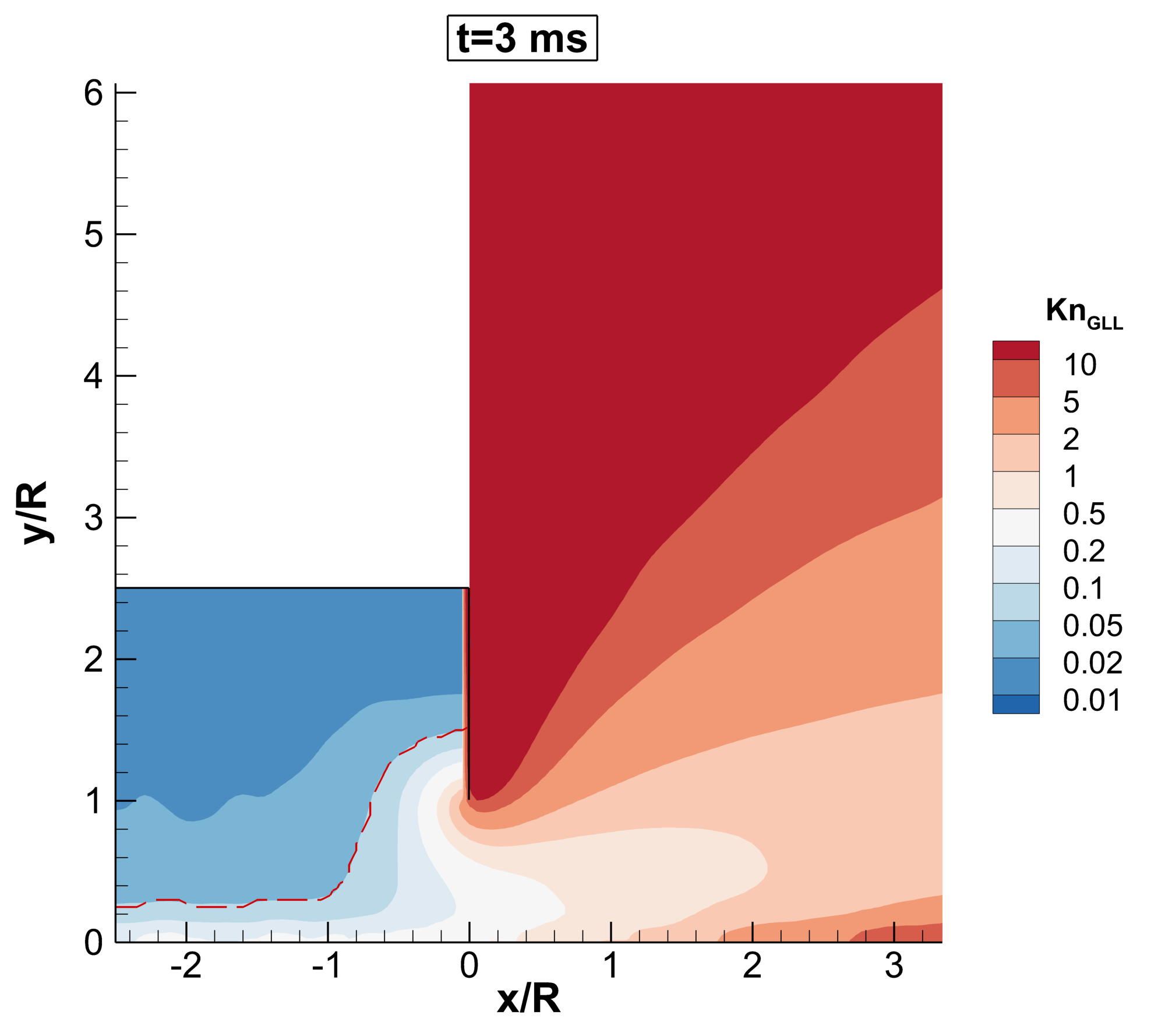}
\includegraphics*[width=0.45\textwidth, keepaspectratio=true]{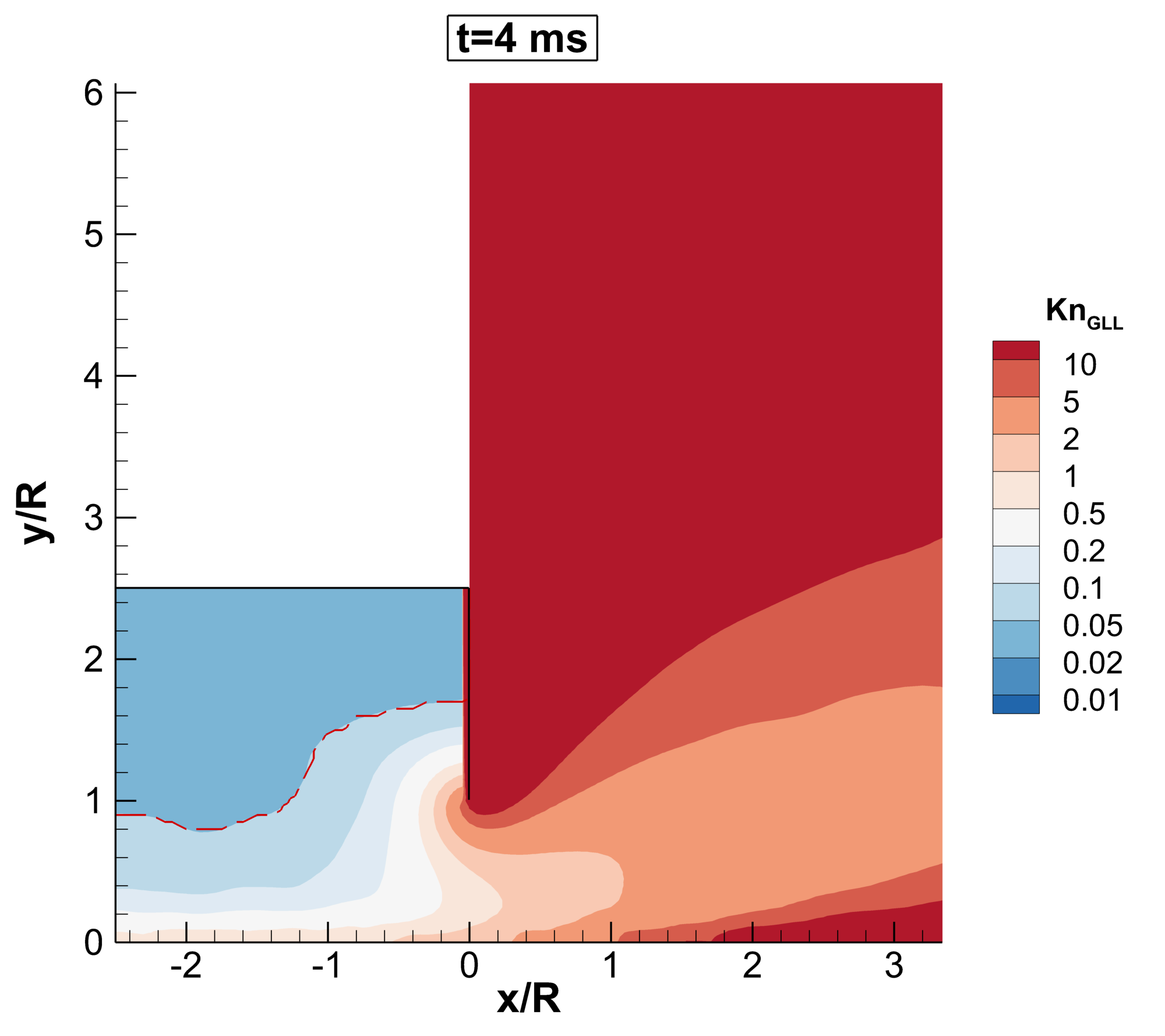}}
}
\caption{Temporal evolution of maximum gradient local length Knudsen number $\mathrm{Kn}_{GLL}$, along with continuum-rarefied interface predicted by uniGasFoam for the transient expansion into vacuum case.}
\label{fig:expansionSliceKn}
\end{figure}

In Fig.~\ref{fig:expansionSliceKn} the temporal evolution of the maximum gradient local length Knudsen number $\mathrm{Kn}_{GLL}$ and continuum-rarefied interface predicted by uniGasFoam is presented. Initially, the rarefied region is mostly concentrated downstream from the vessel separating wall, due to the high gradients and low pressures in this region. As the flow progresses, the gas evacuates the vessel thus decreasing the vessel pressure, and the continuum-rarefied interface shifts downstream, slowly encompassing the vessel region. In this case, the density decomposition criterion is the dominant one, followed by the velocity and temperature criteria. Moreover, the continuum-rarefied interface always closely tracks the isoline $\mathrm{Kn}_{GLL}$=0.05.

\begin{figure}[]
\centerline{
\includegraphics*[width=0.45\textwidth, keepaspectratio=true]{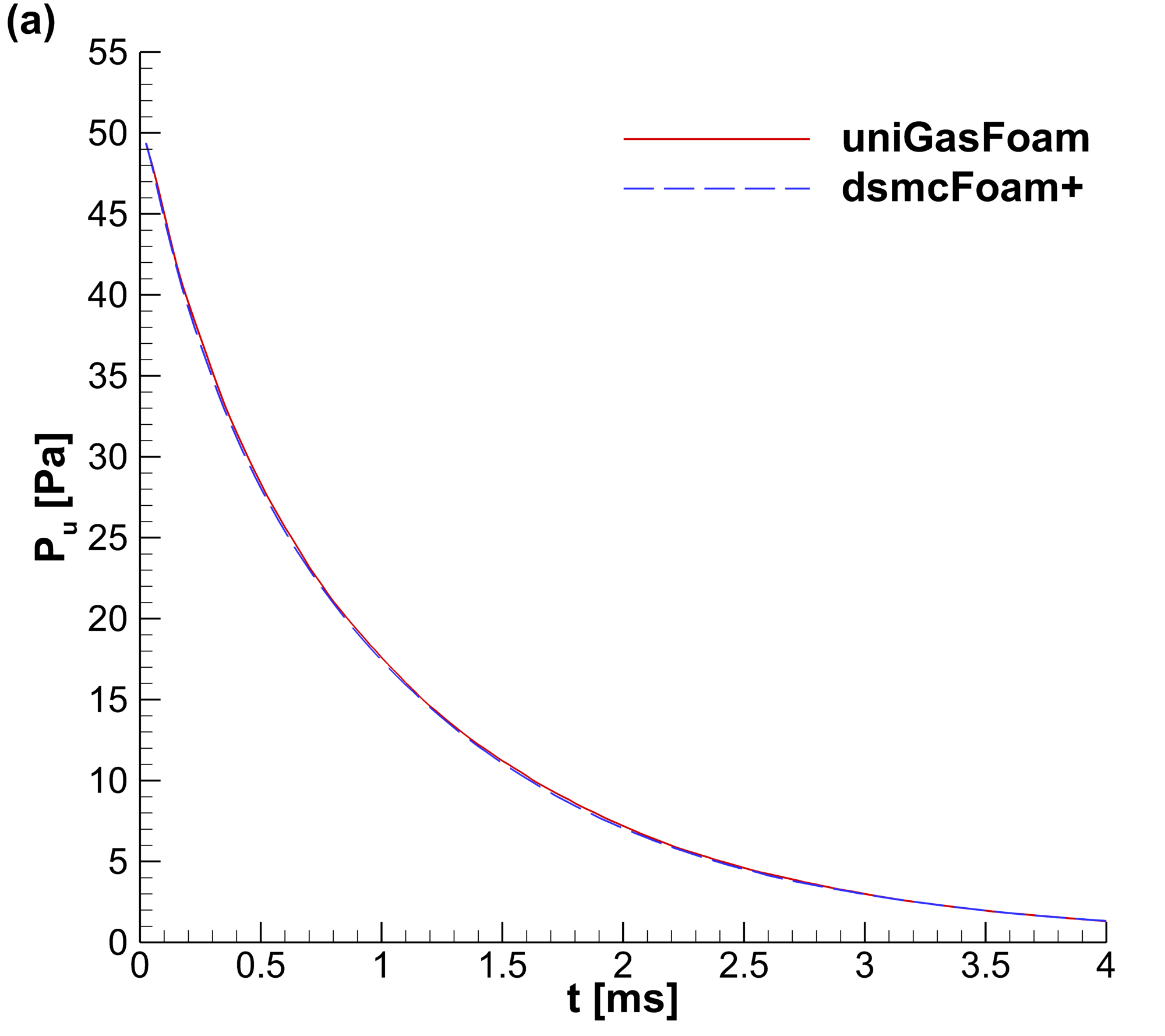}
\includegraphics*[width=0.45\textwidth, keepaspectratio=true]{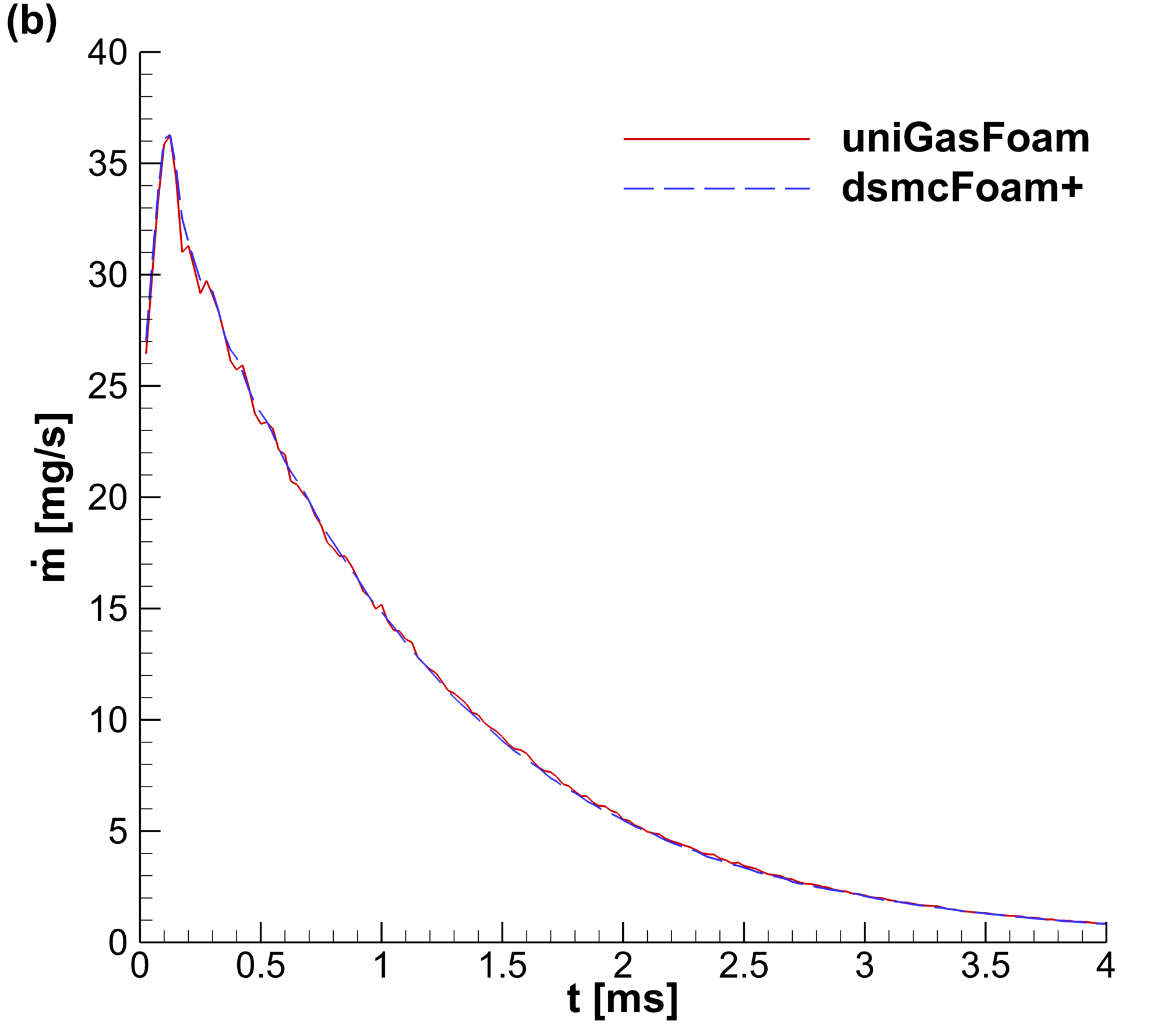}
}
\caption{Temporal evolution of (a) average upstream vessel pressure and (b) orifice mass flow rate obtained by uniGasFoam and dsmcFoam+ for the transient expansion into vacuum case.}
\label{fig:expansionPlot}
\end{figure}

The time-dependent evolution of the average vessel pressure and orifice mass flow rate computed by uniGasFoam and dsmcFoam+ is provided in Fig.~\ref{fig:expansionPlot}. As expected the vessel pressure decreases exponentially, while the mass flow rate initially increases reaching a peak, and then it also decreases exponentially. It is clear that uniGasFoam is in excellent agreement with dsmcFoam+ being able to capture the transient non-equilibrium phenomena occurring during the gas expansion process.

\begin{figure}[]
\centerline{
\includegraphics*[width=0.45\textwidth, keepaspectratio=true]{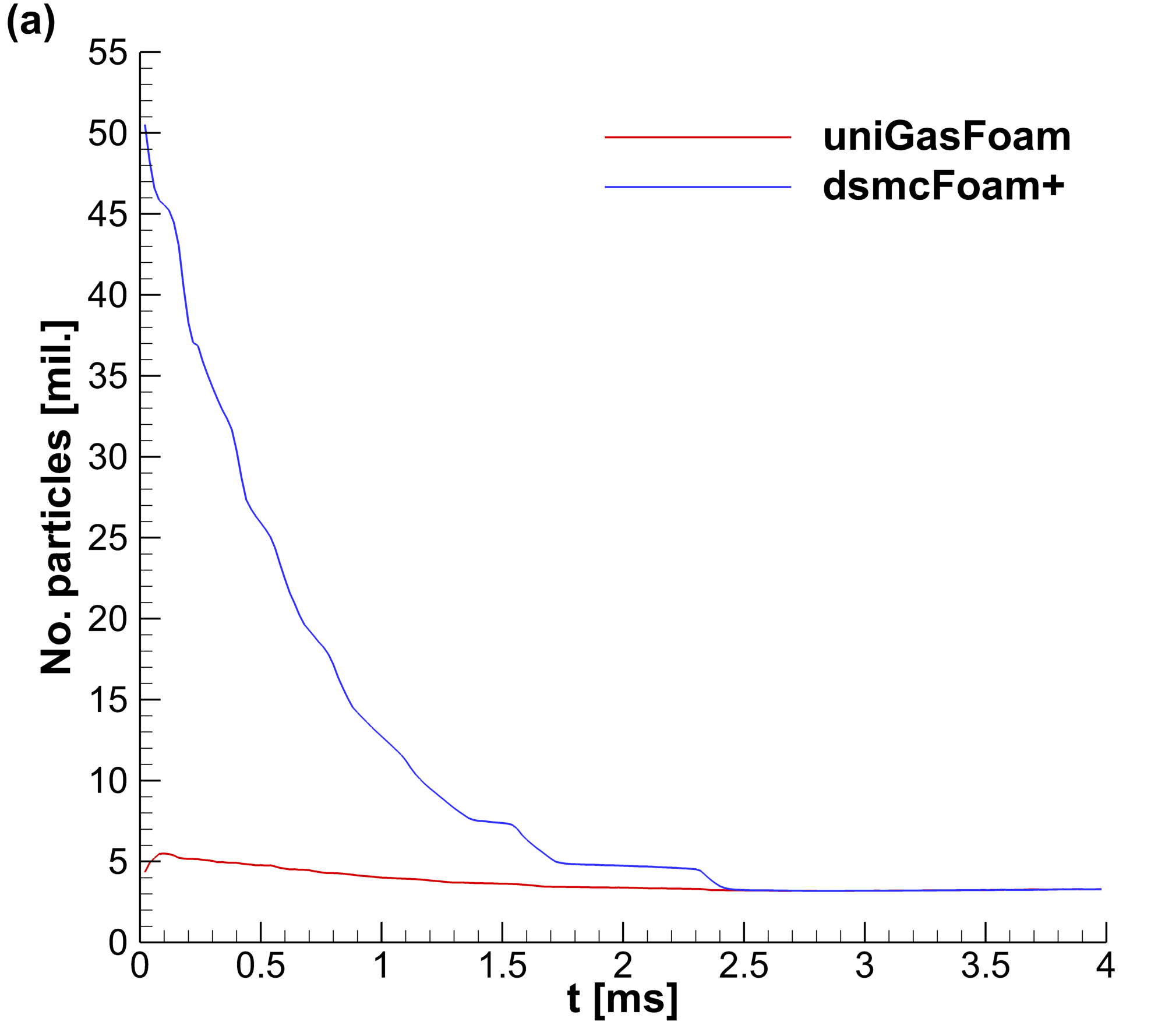}
\includegraphics*[width=0.45\textwidth, keepaspectratio=true]{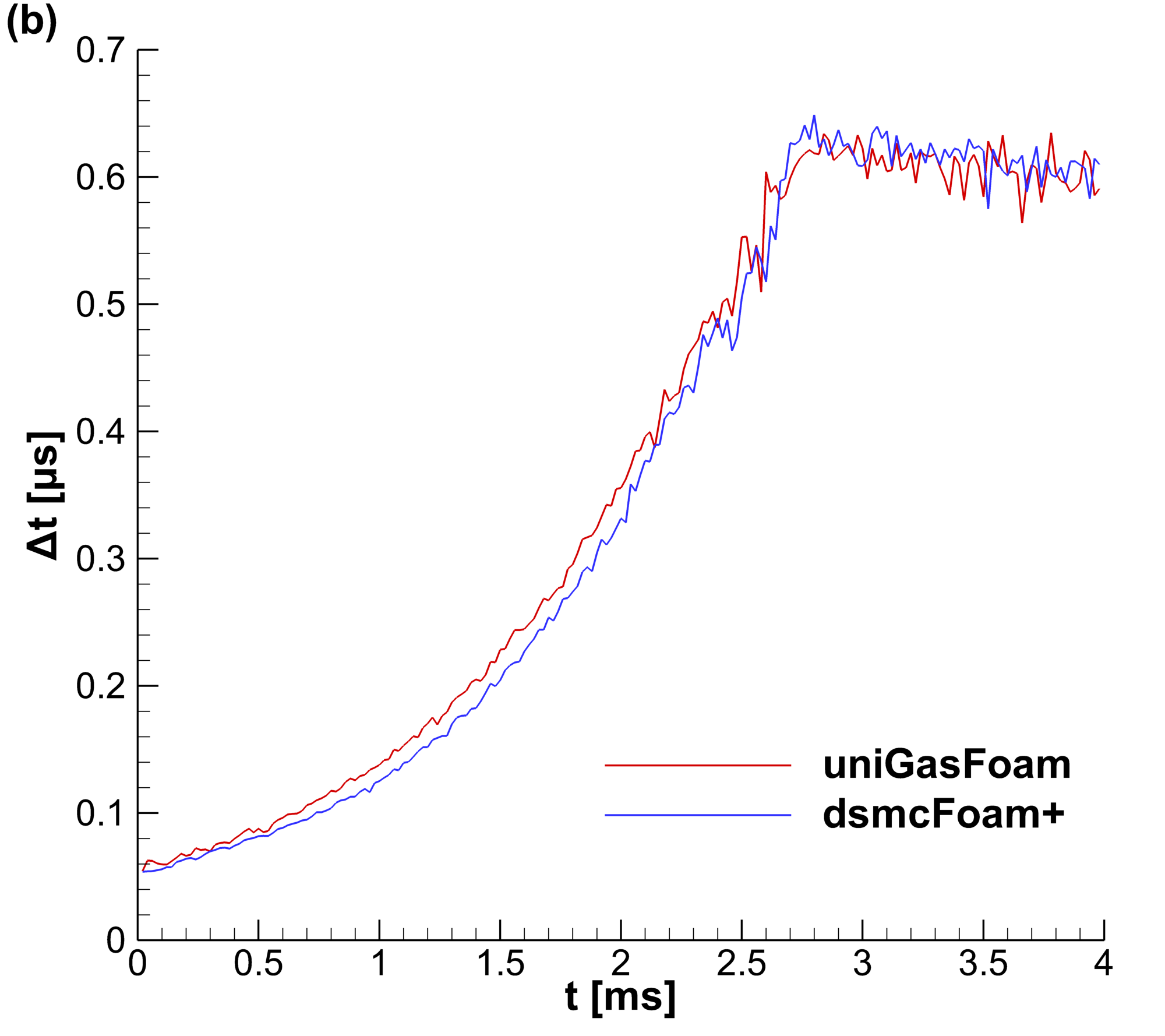}
}
\caption{Temporal evolution of (a) computational particles and (b) time step obtained by uniGasFoam and dsmcFoam+ for the transient expansion into vacuum case.}
\label{fig:expansionPlotCC}
\end{figure}

Focusing on the computational cost, the number of particles and time step employed by dsmcFoam+ and uniGasFoam is illustrated in Fig.~\ref{fig:expansionPlotCC}. The number of particles required by uniGasFoam remains relatively constant throughout the simulation, whereas dsmcFoam+ initially requires approximately ten times more computational particles. As the flow evolves and the rarefied region extends over most of the flow domain, the two solvers converge in terms of required computational particles. Initially, the time step of both solvers is constrained by the vessel's mean collision time. As the flow progresses and the vessel pressure drops, the mean collision time and consequently the maximum allowed time step is automatically increased. At t$\approx$2.7 ms the mean collision time has increased sufficiently to activate the CFL criterion, thereby limiting the maximum allowable time step for the remainder of the simulation. Initially, uniGasFoam implements only a marginally larger time step compared to dsmcFoam+ until both solvers converge in terms of the time step employed. The computational time needed for dsmcFoam+ amounted to 32 hours, using 20 cores. In contrast, uniGasFoam completed the simulation in only 4.6 hours by also employing 20 cores, resulting in an overall speedup of 7.

\section{Conclusions and future work} \label{Sect:conclusions}

A new open-source solver for particle-based multiscale gas flow simulations, uniGasFoam, has been presented. Developed entirely within the OpenFOAM software framework, uniGasFoam builds on the established dsmcFoam+ solver and offers versatile simulation capabilities, including hybrid SP-DSMC and USP-DSMC methods, as well as pure DSMC, SP, and USP gas flow simulations. Based on the presented study, the hybrid USP-DSMC module of uniGasFoam is shown to be the best in terms of performance and accuracy for all multiscale rarefied gas flow simulations.

The solver implements arbitrary 3D geometries and supports parallel simulations using the MPI domain decomposition approach available in OpenFOAM. In addition, sophisticated techniques, such as transient adaptive sub-cells, non-uniform cell weighting and adaptive global time stepping have been introduced in uniGasFoam, to optimise the computational cost with minimal user input, simplifying the usage of the solver and reduce potential sources of errors in the simulation results.

In this study, the USP-DSMC module of uniGasFoam was validated for four benchmark cases, namely the flow past a cylinder, flow over a plate, nozzle plume impingement, and transient expansion into vacuum. In all investigated benchmarks uniGasFoam yielded an excellent agreement with pure DSMC (dsmcFoam+), conventional CFD-DSMC (hybridDCFoam), and literature results. Depending on the specific case, uniGasFoam achieved speedups ranging from 1.3 to 43 compared to dsmcFoam+. The hybrid USP-DSMC approach showed varying degrees of speedup over pure DSMC simulations, with performance gains depending on the number of particles and the time step used. Additionally, it has been shown that in cases where the rarefied region is comparable to or larger than the continuum one, the USP-DSMC scheme outperforms the CFD-DSMC approach in terms of computational speed. Hence, the use of uniGasFoam is justified in multiscale flows where a large range of gas densities and/or dimension scales are present.

Future development of uniGasFoam will focus on extending its capabilities to include polyatomic gases with rotational and vibrational energy modes, as well as multi-species gas mixtures, further enhancing its versatility for complex gas flow simulations.

While further developments are possible, the software presented demonstrates significant potential for simulating multiscale rarefied gas flows in engineering applications where pure DSMC simulations are computationally prohibitive.

\section*{Acknowledgements} \label{sec:acknowledgements}
The authors thank the UK's Engineering and Physical Sciences Research Council (EPSRC) for funding under grant no. EP/\-V012002/\-1, EP/\-V01207X/\-1, and EP/\-V012010/\-1. For the purpose of open access, the authors have applied a CC BY public copyright license to any Author Manuscript version arising from this submission. This work used the ARCHER2 UK National Supercomputing Service (\url{https://www.archer2.ac.uk}).

\bibliographystyle{elsarticle-num}
\bibliography{bibfile.bib}

\end{document}